\documentclass[jor, amsmath, amssymb, preprint, floatfix, unsortedaddress
]{revtex4-2}

\usepackage{graphicx}
\usepackage{dcolumn}
\usepackage{bm}

\usepackage[utf8]{inputenc}
\usepackage[T1]{fontenc}
\usepackage{mathptmx}
\usepackage{etoolbox}

\usepackage{multirow}
\usepackage{enumerate}

\usepackage[dvipsnames]{xcolor}

\newcommand{\Gp}{G^{\prime}(\omega)}
\newcommand{\Gpp}{G^{\prime\prime}(\omega)}
\newcommand{\Tp}{G_{33}^{\prime}}
\newcommand{\Tpp}{G_{33}^{\prime\prime}}

\newcommand{\Gst}{G^{*}(\omega)}
\newcommand{\Tst}{G^{*}_{33}(\omega)}

\begin{document}


\title[Kramers-Kronig for MAOS]{Kramers-Kronig Relations for Nonlinear Rheology: 2. Validation of Medium Amplitude Oscillatory Shear (MAOS) Measurements}

\author{Sachin Shanbhag}%
 \email{sshanbhag@fsu.edu}
\affiliation{Department of Scientific Computing, Florida State University, Tallahassee, FL 32306. USA}

\author{Yogesh M. Joshi}%
\email{joshi@iitk.ac.in}
\affiliation{Department of Chemical Engineering, Indian Institute of Technology, Kanpur, INDIA}

\date{\today}

\begin{abstract}

The frequency dependence of third-harmonic medium amplitude oscillatory shear (MAOS) modulus $G_{33}^{*}(\omega)$ provides insight into material behavior and microstructure in the asymptotically nonlinear regime. Motivated by the difficulty in the measurement of MAOS moduli, we propose a test for data validation based on nonlinear Kramers-Kronig relations. We extend the approach used to assess the consistency of linear viscoelastic data by expressing the real and imaginary parts of $G_{33}^{*}(\omega)$ as a linear combination of Maxwell elements: the functional form for the MAOS kernels is inspired by time-strain separability (TSS). We propose a statistical fitting technique called the SMEL test, which works well on a broad range of materials and models including those that do not obey TSS. It successfully copes with experimental data that are noisy, or confined to a limited frequency range. When  Maxwell modes obtained from the SMEL test are used to predict the first-harmonic MAOS modulus $G_{31}^{*}$, it is possible to identify the range of timescales over which a material exhibits TSS. 
\end{abstract}

\keywords{principle of causality \and regularized least squares \and MAOS \and data validation}

\maketitle

\section{Introduction}

Due to their convenience, oscillatory shear tests have become increasingly important tools for characterizing the linear and nonlinear rheology of soft materials \cite{TschoeglPhenomenological, Ferry1980, Cho2016}. In strain-controlled experiments a sinusoidal strain $\gamma(t) = \gamma_0 \sin \omega t$ with amplitude $\gamma_0$ and angular frequency $\omega$ is applied, and the resulting stress profile $\sigma(t)$ is measured. In the linear viscoelastic (LVE) regime, deformations are infinitesimal so that the equilibrium microstructure of the material is not disturbed. In practice, LVE properties are measured by \textit{small amplitude} oscillatory shear (SAOS) experiments in which $\gamma_0 \approx \mathcal{O}(0.01)$ is small, and the resulting stress response is linear  in $\gamma_0$,
\begin{equation}
\sigma_\text{SAOS}(t) = \gamma_0 \left( \Gp \sin \omega t + \Gpp \cos \omega t\right),
\label{eqn:stressSAOS}
\end{equation}
where $\Gp$ and $\Gpp$ are the LVE storage and loss moduli, respectively. They are intrinsic material properties that are independent of the strain amplitude, and correspond to the real and imaginary parts of the complex relaxation modulus $\Gst = \Gp + i \Gpp$. The principle of causality  induces a mathematical relationship between the real and imaginary parts of $\Gst$ called the Kramers-Kronig relations (KKR) \cite{L.Kronig1926, Kramers1927}. For viscoelastic liquids,
\begin{align}
	G^{\prime}(\omega) & = -\dfrac{2 \omega^2}{\pi} \int_{0}^{\infty} \dfrac{G^{\prime\prime}(u)/u}{u^2 -  \omega^2} du \notag\\
	G^{\prime\prime}(\omega) & = \dfrac{2 \omega}{\pi} \int_{0}^{\infty}  \dfrac{G^{\prime}(u)}{u^2 - \omega^2} du,
\label{eqn:kk_saos}
\end{align}
Due to the singularity at $u = \omega$ in the denominator, the Cauchy principal value of these integral transforms is automatically implied. We call these relations \textit{linear} KKR, since they correspond to the LVE moduli. They can be expressed in different forms (see Table 1 of companion paper ref. \cite{kkr1}).  Besides rheology \cite{TschoeglPhenomenological}, equivalent linear KKR find use in numerous other areas including optics \cite{Peiponen1991, Lucarini2005}, electrochemical impedance spectroscopy \cite{Gross1941, Boukamp2004}, electrical networks \cite{Bode1945}, etc.

\subsection{Applications of Linear Kramers-Kronig Relations}

Linear KKR are used to either  \textit{numerically evaluate} one signal from the other, or to \textit{test the consistency} of the two signals. In the first scenario, KKR is used to compute either the real or imaginary component ($\Gp$ or $\Gpp$) from the other. This is useful when one of the signals is weak, and perhaps falls below the limits of instrument sensitivity. This is also the common setting in optics, where it is easier to measure the imaginary part (absorption coefficient) of the complex dielectric permittivity over a broad range of frequencies, and infer the real part (refractive index) by numerically integrating the KKR \cite{Lucarini2005}. 

Even the earliest attempts to numerically integrate KKR recognized the need to deal with the potential singularity at $u = \omega$, and the extrapolation of experimental observations beyond the finite frequency window, $\omega_{\min} \leq \omega \leq \omega_{\max}$, over which they are measured \cite{Silva1941}. Since the singularity is an integrable Cauchy-type singularity, standard methods like linearization \cite{Silva1941, Lovell1974}, or integration by parts may be used \cite{Davis1984, Amari1995,  Urquidi-Macdonald1986, Urquidi-Macdonald1990}. Custom Gauss quadrature methods have also been developed to specifically tackle this problem \cite{King2002, King2007}.  Since the integral in the KKR extends from zero to infinity, it is preferable to obtain data over the widest possible experimental window $[\omega_{\min}, \omega_{\max}]$. Nevertheless, the question of how to optimally extend the data on both ends remains. Different functional forms including polynomials \cite{Urquidi-Macdonald1990}, polynomial functions of the logarithm of the frequency \cite{Esteban1991}, and splines \cite{Bakry2018}, have been previously employed for extrapolation.

In the second scenario, KKR are used for \textit{data validation}, where the consistency of the measured signals, $\Gp$ and $\Gpp$, is evaluated. This is practically relevant, for example, when time-temperature superposition is used to construct master-curves by shifting a number of individual datasets \cite{Rouleau2013}. Sometimes, an independent parameter like strain rate, stress, pH, etc. is observed to play a role similar to temperature in time-temperature superposition. In such situations, KKR becomes a useful tool to check the authenticity of the superposition of the experimental data. This strategy was successfully used, for example, by Erwin and coworkers \cite{Erwin2010} to determine the validity of strain-rate frequency superposition in soft materials \cite{Wyss2007}. Numerical evaluation of the KKR integrals is one method to check consistency: we can compare the experimental $\Gp$ and $\Gpp$ with the moduli calculated from KKR (equation \ref{eqn:kk_saos}).

However, if the goal is merely to test the consistency of the experimental data and KKR, a simple strategy that avoids the problems associated with numerical integration can be adopted. In this approach, we attempt to infer a relaxation spectrum by simultaneously fitting $\Gp$ and $\Gpp$ to a set of $N$ discrete Maxwell modes $\mathcal{M} = \{g_j, \tau_j\}$ with $j = 1, \cdots, N$, which is called the discrete relaxation spectrum (DRS) \cite{Winter1997, Boukamp1995, Agarwal1992},
\begin{align}
\Gp &  = \sum_{j=1}^{N} g_j \dfrac{\omega^2 \tau_j^2}{1 + \omega^2 \tau_j^2} = \sum_{j=1}^{N} g_j k^{\prime}(\omega \tau_j) \notag\\
\Gpp & = \sum_{j=1}^{N} g_j \dfrac{\omega \tau_j}{1 + \omega^2 \tau_j^2} = \sum_{j=1}^{N} g_j k^{\prime \prime}(\omega \tau_j),
\label{eqn:drs}
\end{align}
where $g_j > 0$ and $\tau_j > 0$ are the modulus and timescale characterizing the $j^\text{th}$ relaxation mode, respectively \cite{TschoeglPhenomenological}. $k^{\prime}(z) = z^2/(1+z^2)$ and $k^{\prime\prime}(z) = z/(1+z^2)$ are the kernels corresponding to the storage and loss moduli, respectively. The DRS can be inferred from frequency sweep experiments using open-source programs like DISCRETE \cite{provencher76}, pyReSpect \cite{Takeh2013, Shanbhag2019respect, Shanbhag2020}, or a commercial program like IRIS \cite{baumgaertel89a}. While the method proposed later in this work is inspired by the DRS, it should be noted that the continuous analogue of equation \ref{eqn:drs}, called the continuous relaxation spectrum (CRS), where the summation is replaced by an integral may also used in lieu of the DRS for data validation. 

The attractive feature of this approach is that KKR are built into the kernel functions by design: $k^{\prime}$ and $k^{\prime\prime}$ obey equation \ref{eqn:kk_saos}. Therefore, any linear combination such as that considered in the DRS (equation \ref{eqn:drs}) necessarily obeys KKR. If a DRS that can simultaneously account for the storage and loss moduli \textit{cannot} be found, the validity of the experimental data is brought into question, because it violates KKR. Since this method relies on optimization as its engine, it is well-suited to experimental data that are noisy, or confined to a limited frequency window.

\subsection{Kramers-Kronig Relations for Medium Amplitude Oscillatory Shear}

In medium amplitude oscillatory shear (MAOS) tests, we impose a sinusoidal strain $\gamma(t) = \gamma_0 \sin \omega t$, similar to SAOS measurements. As $\gamma_0$ is gradually increased, the weakest or asymptotically nonlinear modes are initially activated \cite{Hyun2007, Wagner2011, Hyun2011, Ewoldt2013, Bharadwaj2015, Cho2016}. Due to the odd symmetry of shear stress with shear strain, these weak nonlinear modes are proportional to $\gamma_0^3$. In this regime, the total stress is given by, $\sigma(t) =  \sigma_\text{SAOS}(t) +  \sigma_\text{MAOS}(t) + \mathcal{O}(\gamma_0^5)$, where \cite{Pearson1982},
\begin{equation}
\sigma_\text{MAOS}(t) = \gamma_0^3 \left[G_{31}^{\prime}(\omega) \sin \omega t +  G_{31}^{\prime\prime}(\omega) \cos \omega t +G_{33}^{\prime}(\omega) \sin 3\omega t +  G_{33}^{\prime\prime}(\omega) \cos 3\omega t \right]
\label{eqn:sigmaMAOS}
\end{equation}
The MAOS moduli associated with the first and third harmonic are $G^{*}_{31} = G^{\prime}_{31} + i G^{\prime\prime}_{31}$, and $G^{*}_{33} = G^{\prime}_{33} + i G^{\prime\prime}_{33}$, respectively. These MAOS moduli, primarily the third-harmonic $\Tst$, have been used extensively to probe features of material structure that are not salient in LVE data. For example the intrinsic nonlinearity parameter $Q_0$, which is related to the relative intensity of the third harmonic normalized by the first
harmonic $I_3/I_1=I_{3/1}(\omega, \gamma_0)$ as \cite{Hyun2009, Wilhelm2002, Cho2016},
\begin{equation}
Q_0(\omega) = \lim_{\gamma_0 \rightarrow 0} \dfrac{I_{3/1}(\omega, \gamma_0)}{\gamma_0^2} = \dfrac{|\Tst|}{|\Gst|}.
\label{eqn:Q0}
\end{equation}
The shape of $Q_0(\omega)$ curves is sensitive to polymer architecture. Therefore, it can be used to distinguish between linear and branched polymers \cite{Hyun2009, Wagner2011, Song2016} by analyzing the number of local peaks. $Q_0(\omega)$ is also more sensitive to effects of fillers in polymer nanocomposites, and can be used to evaluate nanoparticle dispersion quality \cite{Lee2016, Lim2013}, and droplet size dispersion in polymer blends \cite{Ock2016, Salehiyan2014}, etc. In contrast, the first-harmonic MAOS moduli $G_{31}^{*}(\omega)$ have been less frequently used \cite{Song2019, Xiong2018, Carey-DeLaTorre2018}.

Using a multiple integration formulation to capture the nonlinear mechanical response, and appealing to the principle of causality, KKR can be derived for $\Tst$ \cite{kkr1}. It can be succinctly represented in complex notation as,
\begin{equation}
G_{33}^{*}(\omega) = \dfrac{i}{\pi} \omega^3 \int_{-\infty}^{\infty} \dfrac{1}{u^3} \dfrac{G_{33}^{*}(u)}{u - \omega} du.
\label{eqn:kkr_G33}
\end{equation}
Similar to the linear KKR (equation \ref{eqn:kk_saos}), this relation can be expressed as a pair of equations relating the real and imaginary parts of $G_{3}^{*}$ on a non-negative frequency domain,
\begin{align}
G_{33}^{\prime}(\omega) & = -\dfrac{2\omega^4}{\pi} \int_{0}^{\infty} \dfrac{G_{33}^{\prime \prime}(u)/u^3}{u^2 - \omega^2} du \notag\\
G_{33}^{\prime\prime}(\omega) & = \dfrac{2 \omega^3}{\pi} \int_{0}^{\infty} \dfrac{G_{33}^{\prime}(u)/u^2}{u^2 - \omega^2} d u,	\label{eqn:kk_maos}
\end{align}
Just like linear KKR, these MAOS KKR can be used to numerically evaluate one signal from the other, or for data validation. Analogous KKR are widely used in nonlinear optics \cite{Hutchings1992, Peiponen2004, Boyd2008}. As described in the ref. \cite{kkr1}, KKR for $G_{31}^{*}(\omega)$ do not exist, and there is no direct way in which their consistency can be similarly evaluated. However, MAOS functions are related to $\Gst$ for materials exhibiting time-strain separability, i.e. when the nonlinear shear relaxation modulus $G(t, \gamma) = h(\gamma) G(t)$ in step-strain experiments. Here, $h(\gamma)$ is the damping function, and $G(t)$ is the LVE stress relaxation modulus. These functions are given by \cite{Pearson1982, Cho2010, Martinetti2019, Lennon2020, Liu2020}:
\begin{align}
G_{31}^{\prime}(\omega)& = \dfrac{3a}{4} \left[4 \Gp - G^{\prime}(2\omega) \right] \notag \\
G_{31}^{\prime\prime}(\omega) & = \dfrac{3a}{4} \left[ 2 \Gpp - G^{\prime\prime}(2\omega) \right] \notag \\
G_{33}^{\prime}(\omega) & = -\dfrac{a}{4} \left[3 \Gp - 3 G^{\prime}(2\omega) + G^{\prime}(3\omega) \right] \notag\\
G_{33}^{\prime\prime}(\omega) & = -\dfrac{a}{4} \left[3 \Gpp - 3 G^{\prime\prime}(2\omega) +  G^{\prime\prime}(3\omega) \right],
\label{eqn:MAOS_G3}
\end{align}
where $a  = \lim_{\gamma \rightarrow 0} dh/d(\gamma^2)$ is the derivative of the damping function in the limit of zero shear. 

\subsection{Motivation and Scope}

Obtaining MAOS moduli experimentally is tedious, and fraught with numerous potential sources of error. This is in contrast to the ease with which LVE moduli $\Gp$ and $\Gpp$ can be obtained. This difficulty arises from different sources: (i) the window of suitable strain amplitudes $\gamma_0$ is narrow \cite{Ewoldt2013}. MAOS signals are weak when $\gamma_0$ is small, and contaminated by the higher-harmonics when $\gamma_0$ is too large. (ii) the optimal $\gamma_0$ is frequency-dependent: at low frequency, larger strain amplitudes are desirable \cite{Singh2018}. (iii) the method is indirect: the ``true'' MAOS moduli are extracted by extrapolating measurements at multiple strain-amplitudes to filter out the contribution of higher harmonics.

This is a complicated process, which only serves to increase the importance of data validation. As mentioned previously, consistency of LVE moduli with linear KKR can be tested by fitting a DRS. In this work, we adopt a similar approach, and devise an efficient test to quantitatively assess the consistency of MAOS moduli $G_{33}^{*}$; fortunately a majority of practically used MAOS tests involve $G_{33}^{*}$.

The proposed method involves fitting $\Tp$ and $\Tpp$ to a sum of Maxwell elements as discussed in section \ref{sec:methods}. The fitting is performed using a statistical technique called LASSO (least absolute shrinkage and selection operator) regression \cite{Tibshirani1996, Tibshirani2011}. It is a regularized linear least-squares regression method that automatically selects a parsimonious set of modes. We call this technique \underline{s}um of \underline{M}axwell \underline{e}lements using \underline{L}ASSO, or \textit{SMEL test} after the underlined initials. The SMEL test is applied to the MAOS response of the Giesekus model, which is not time-strain separable (TSS), a TSS power-law model, and experimental data on a well-characterized solution of Polyvinyl alcohol (PVA) and Borax \cite{Ewoldt2013, Bharadwaj2015, bharadwaj2016dissertation}.

\section{Methods For Data Validation}
\label{sec:methods}

We assume that MAOS experimental data for the third-harmonic modulus $G_{33}^{*}(\omega) = \Tp(\omega) + i \Tpp(\omega)$ are available over a finite frequency window $\omega \in [\omega_{\min}, \omega_{\max}]$ as $\mathcal{D} = \{\omega_i, D^{\prime}_i = \Tp(\omega_i), D^{\prime\prime}_i = \Tpp(\omega_i) \}$ for $i = 1, 2, \cdots, n_d$. Here, $n_d$ is the number of data-points, $\omega_1 = \omega_{\min}$, and $\omega_{n_d} = \omega_{\max}$. Typically, but not necessarily, these data are evenly spaced on a logarithmic frequency scale. Our goal in this section is to develop an efficient method to test whether these observations violate the MAOS KKR given by equation \ref{eqn:kk_maos}.

\subsection{Functional Form for Kernel}

The first step is to devise a suitable functional form for the MAOS kernels, $K^{\prime}$ and $K^{\prime\prime}$, that mimics the relationship between the linear kernels ($k^{\prime}$ and $k^{\prime\prime}$ in equation \ref{eqn:drs}) and the linear KKR (equation \ref{eqn:kk_saos}). This functional form should (i) automatically satisfy the MAOS KKR (equation \ref{eqn:kk_maos}), and (ii) be flexible enough to assimilate the behavior of a wide class of materials. As with the DRS, the second requirement can potentially be addressed by considering a large set of modes, $N$.

To satisfy the first requirement, we appeal to the MAOS response of a single TSS Mawell mode with relaxation time $\tau$. Letting $z = \omega \tau$, the LVE kernels are $k^{\prime}(z) = z^2/(1 + z^2)$ and $k^{\prime\prime}(z) = z/(1 + z^2)$. The MAOS functions $\Tp$ and $\Tpp$ are related to the LVE moduli $\Gp$ and $\Gpp$ and the damping function via equation \ref{eqn:MAOS_G3} \cite{Pearson1982, Cho2010, Martinetti2019, Lennon2020, Liu2020}. Using this relationship, we propose the MAOS kernels,
\begin{align}
K^{\prime}(z) & = 3 k^{\prime}(z) - 3 k^{\prime}(2z) + k^{\prime}(3z) = \dfrac{36 z^4 (z^2 - 1)}{(1+z^2)(1+4z^2)(1+9z^2)}\notag\\
K^{\prime\prime}(z) & = 3 k^{\prime\prime}(z) - 3 k^{\prime\prime}(2z) +  k^{\prime\prime}(3z) = \dfrac{6z^3(11z^2 - 1)}{(1+z^2)(1+4z^2)(1+9z^2)}.
\label{eqn:MAOS_kernel}
\end{align}
These MAOS kernels automatically satisfy the corresponding MAOS KKR for $G_{33}^{*}$ (equation \ref{eqn:kk_maos}):
\begin{align}
K^{\prime}(\omega) & = -\dfrac{2\omega^4}{\pi} \int_{0}^{\infty} \dfrac{K^{\prime \prime}(u)/u^3}{u^2 - \omega^2} du \notag\\
K^{\prime\prime}(\omega) & = \dfrac{2 \omega^3}{\pi} \int_{0}^{\infty} \dfrac{K^{\prime}(u)/u^2}{u^2 - \omega^2} d u.	\label{eqn:kk3_kernel}
\end{align}
Thus, the idea is to fit the experimental data $\mathcal{D}$ to a linear combination of MAOS kernels, so that
\begin{align}
G_{33}^{\prime}(\omega) & \approx P^{\prime}(\omega) = \sum_{j=1}^{N} g_j K^{\prime}(\omega \tau_j) \notag\\
G_{33}^{\prime\prime}(\omega) & \approx P^{\prime\prime}(\omega) = \sum_{j=1}^{N} g_j K^{\prime \prime}(\omega \tau_j),
\label{eqn:drs3}
\end{align}
where $g_j$ and $\tau_j$ are the weights and timescales associated with the $j^\text{th}$ mode. They can be determined by fitting experimental data at $\omega = \omega_1, \cdots, \omega_{n_d}$ using equation \ref{eqn:drs3}. Once the modes $\mathcal{M} = \{g_j, \tau_j\}$ are found, they can be used to predict the corresponding MAOS response, $P^{\prime}$ and $P^{\prime\prime}$, at any frequency. Since $K^{\prime}$ and $K^{\prime\prime}$ satisfy KKR by design, the validity of the experimental data can be ascertained by the examining the quality of the fit.

\subsection{Numerical Method for Fitting MAOS modes}

Now that the framework for data validation is established, we turn our attention to the numerical method for fitting the experimental data with a set of kernel functions.  We formulate a weighted least-squares problem by defining the objective function via a sum of squared residuals (SSR),
\begin{equation}
\chi^2 = \dfrac{1}{4 n_d} \sum_{i=1}^{n_d}  \left[w_i^{\prime} \left(D_i^{\prime} - P^{\prime}(\omega_i) \right)^2 + w_i^{\prime\prime} \left(D_i^{\prime\prime} - P^{\prime\prime}(\omega_i) \right)^2 \right],
\label{eqn:chi}
\end{equation}
where the terms inside the two parentheses in the square brackets are residuals at a particular frequency $\omega_i$. The positive weights $w_i^{\prime} = 1/|D^{\prime}_i|$ and $w_i^{\prime\prime} = 1/|D^{\prime\prime}_i|$ are chosen to prevent the contribution of small $D_i^{\prime}$ and $D_i^{\prime\prime}$ from being overwhelmed. This improves the agreement between the data and the fit, when the moduli are presented on a log-log plot.

For a given number of modes $N$, the least-squares problem involves finding the optimal set of modes $\mathcal{M}$ that minimizes the objective function $\chi^2$. In general, this is a nonlinear least-squares problem that requires a sophisticated approach \cite{Shanbhag2020}, especially when the number of modes $N$ is unknown, and a parsimonious representation is desired. Sometimes, values of $\tau_j$ are pre-specified on a regular logarithmically spaced grid. This automatically fixes $N$, and significantly simplifies the problem: (i) the number of parameters to determine is halved from $2N$ to $N$, and (ii) the regression problem becomes linear in the undetermined coefficients $g_j$. 

Furthermore, unlike the DRS, we do not require these coefficients $g_j$ to be positive: the MAOS kernels (equation \ref{eqn:kk3_kernel}) and their linear combinations (equation \ref{eqn:drs3}) can simply be viewed as a means to an end (data validation), and not objects of interest themselves. This further simplifies the problem. Despite these simplications, the resulting \textit{linear} least squares problem is ill-conditioned, which makes it susceptible to noise in the experimental data and round-off errors. We are faced with a common tradeoff: using a large number of modes enhances flexibility, but simultaneously worsens the conditioning of the problem. A standard approach to mitigate this problem is regularization, which improves the conditioning of the problem by adding a constraint to the objective function. 

Here, we consider a regularized linear regression technique called LASSO (least absolute shrinkage and selection operator) \cite{Tibshirani1996, Tibshirani2011}. It modifies the least-squares objective function $\chi^2$ by appending an $L_1$ regularization term as,
\begin{equation}
\chi^2_\text{LASSO}(\{g_j\}) = \chi^2(\{g_j\}) + \alpha \sum_{j=1}^{N} \left|g_j \right|.
\label{eqn:chilasso}
\end{equation}
The parameter $\alpha$ controls the strength of the regularization constraint. When $\alpha = 0$, we recover the original least-squares problem which is poorly conditioned. As $\alpha$ is increased, regularization kicks in and improves conditioning. However, as $\alpha \rightarrow \infty$, the regularization constraint dominates the solution, and drives it to the trivial solution $g_j = 0$ for all $j$. The optimal value of $\alpha$ lies somewhere between these two limits, and seeks to balance the need to describe the experimental data accurately, and the need to pose a well-conditioned problem. Here, optimal value of $\alpha$ is found by 3-fold cross-validation using the built-in function \texttt{LassoCV} from the \texttt{linear\_model} module of the Python machine learning library \texttt{scikit.learn} version 1.01 \cite{Pedregosa2011}. This implementation of LASSO uses coordinate descent to fit the unknown coefficients, and uses a duality gap calculation to control convergence \cite{Friedman2010, Kim2008}. This method is well-suited when the number of modes $N \gtrsim n_d$. An attractive feature of LASSO is feature selection: it automatically identifies the most important modes, and sets the weights $g_j = 0$ for the other modes \cite{Tibshirani2011}. Thus, it provides a parsimonious representation of the data, which while not necessary, provides some insight into the regressed parameters.

\subsection{SMEL Test Algorithm}
\label{sec:algo}

The kernel functions are specified by equation \ref{eqn:MAOS_kernel}. We seek to fit the experimental data $\mathcal{D}$ to a sum of these modes (equation \ref{eqn:drs3}) by minimizing the regularized objective function (equation \ref{eqn:chilasso}). Here, we specify the algorithm for the proposed method, which incorporates these ideas, and checks compliance of $\Tst$ data with MAOS KKR.

In the description below, vectors and matrices are represented using bold symbols, e.g $\bm{X}$. $\bm{X}_i$ denotes the $i^\text{th}$ element of the vector $\bm{X}$; similarly, $\bm{X}_{i, j}$ denotes the element in the $i^\text{th}$ row, and $j^\text{th}$ column of matrix $\bm{X}$. For consistency and brevity, the index $i = 1, 2, \cdots, n_d$ is exclusively used to mark experimental data-points, and the index $j = 1, 2, \cdots, N$ is exclusively used to mark Maxwell modes throughout this work.

\begin{enumerate}
\item \textbf{Setup Data and Parameters}
\begin{itemize}
\item Collect experimental observations, $\mathcal{D} = \{\omega_i, D_i^{\prime} = \Tp(\omega_i), D_i^{\prime\prime}=\Tpp(\omega_i)\}$. Stack these moduli into a $2n_d \times 1$ column vector $\bm{D}$ so that $\bm{D}_{i} =   D_i^{\prime}$ and $\bm{D}_{n_d + i} =   D_i^{\prime\prime}$;
\item Denote the boundaries of the frequency window $\omega_{\min} = \min\{\omega_i\}$ and $\omega_{\max} = \max\{\omega_i\}$; mark the boundaries of the modes  $\tau_{\min} = 0.1/\omega_{\max}$ and $\tau_{\max} = 10/\omega_{\min}$ by extending the experimental domain by one decade on either side;
\item Set mode density $\rho_N = 10$ modes/decade. Set the number of modes $N = \rho_N \cdot \text{floor}(\log_{10} (\tau_{\max}/\tau_{\min}))$;
\item Set the intermediate timescales $\tau_j$ on a logarithmically equispaced grid via,
\begin{equation}
\dfrac{\tau_j}{\tau_{\min}} = \left(\dfrac{\tau_{\max}}{\tau_{\min}} \right)^{\dfrac{j-1}{N-1}},
\end{equation}
Thus, $\tau_1 = \tau_{\min}$ and $\tau_N = \tau_{\max}$.
\end{itemize}
\item \textbf{Setup for LASSO}
\begin{itemize}
\item Furnish two $n_d \times N$ kernel matrices $\bm{K}^{\prime}_{i,j} = K^{\prime}(\omega_i \tau_j)$, and $\bm{K}^{\prime\prime}_{i,j} = K^{\prime\prime}(\omega_i  \tau_j)$ using equation \ref{eqn:MAOS_kernel}. Stack $\bm{K}^{\prime}$ above $\bm{K}^{\prime\prime}$ to produce the $2n_d \times N$ feature matrix $\bm{K}$, so that $\bm{K}_{i,j} = \bm{K}^{\prime}_{i,j}$ and $\bm{K}_{n_d+i,j} = \bm{K}^{\prime \prime}_{i,j}$;
\item Let $\bm{g} = [g_1, \cdots, g_N]^T$ be a column vector of coefficients to be determined so that $\bm{D} \approx \bm{K g}$ (equation \ref{eqn:drs3});
\item Define a $2n_d \times 2n_d$ diagonal matrix of weights $\bm{W}_{ii} = 1/\sqrt{|\bm{D}_i|}$ for weighted least-squares;
\item Transform the data vector $\bm{D}$ and the feature matrix $\bm{K}$ using these weights, $\hat{\bm{D}} = \bm{W D}$ and $\hat{\bm{K}} = \bm{W K}$. The least-squares objective function (equation \ref{eqn:chi}) can be succinctly represented as,
\begin{equation}
\chi^2 = \dfrac{1}{4n_d} (\hat{\bm{D}} - \hat{\bm{K}} \bm{g})^T (\hat{\bm{D}} - \hat{\bm{K}} \bm{g}).
\label{eqn:chi2_unreg}
\end{equation}
The standard unregularized normal equations are $\hat{\bm{K}}^T \hat{\bm{K}} \bm{g} = \hat{\bm{K}}^T \bm{\hat{D}}$;
\item Use the \texttt{scikit-learn} function \texttt{LassoCV} with three-fold cross-validation to determine an optimal value of $\alpha$ in equation \ref{eqn:chilasso}. Solve and determine the coefficients $\bm{g}$;
\item Assess the quality of the fit using the coefficient of determination, or $R^2$, as a proxy for the quality of the fit. If $R^2 \ge 0.95$ (or some other reasonable threshold), the dataset is deemed consistent with MAOS KKR. Otherwise it is deemed inconsistent.
\end{itemize}

\end{enumerate}

For conveniene, Python code that implements the SMEL test is presented in supplementary material. The implementation takes fewer than 20 lines of code.

\section{Results}

The SMEL test is based on LASSO regression using a sum of Maxwell kernel functions inspired by the MAOS response of a TSS Maxwell model. The applicability and generality of the method needs to be evaluated. We consider two synthetic examples for which analytical forms of $\Tst$  are available: (i) a single mode Giesekus model, which violates TSS, and (ii) a TSS material that exhibits power-law LVE and MAOS behavior over a finite frequency window. Note that the MAOS KKR hold for both TSS and non-TSS materials. Furthermore, power-law dependence is often difficult for discrete Maxwell modes to capture. These examples are designed with this aspect in mind. We also explore how the SMEL test responds when we contaminate synthetic data with noise, or arbitrarily shift one of $\Tp$ or $\Tpp$ to artificially create an invalid dataset. We also consider an experimental dataset on a PVA-borax system studied by Ewoldt and Bharadwaj \cite{bharadwaj2016dissertation,Bharadwaj2015}. Finally, implications for the first harmonic $G_{31}^{*}$ and time-strain superposability are discussed.

\subsection{Giesekus Model}

The Giesekus model is a popular constitutive model for polymer solutions and melts \cite{ Giesekus1982, larsoncf}, and for worm-like micelles \cite{Fischer1997, Helgeson2010, KateGurnon2012}. The polymer contribution to the total stress tensor $\bm{\sigma}$ is given by,
\begin{equation}
 \stackrel{\triangledown}{\bm{\sigma}} + \dfrac{1}{\tau} \bm{\sigma} + \dfrac{\alpha_G}{G\tau}\, \bm{\sigma} \cdot \bm{\sigma} = 2 G \bm{D},
\label{eqn:giesekus}
\end{equation}
where $G$ and $\tau$ are the modulus and relaxation time, respectively. The symmetric deformation gradient tensor $\bm{D}$ can be expressed in terms of the velocity gradient $\bm{\nabla v}$ as $\bm{D} = (\bm{\nabla v} + \bm{\nabla v}^T)/2$. For homogeneous flows, the upper-convected derivative simplifies to,
\begin{equation}
\stackrel{\triangledown}{\bm{\sigma}} = \dfrac{\partial \bm{\sigma}}{\partial t} - \bm{\nabla v}^{T} \cdot \bm{\sigma} - \bm{\sigma} \cdot \bm{\nabla v}.
\label{eqn:uc}
\end{equation}
Nonlinearity is subsumed into a single nonlinear parameter $\alpha_G \in [0, 1]$. When $\alpha_G = 0$, it is equivalent to the upper-convected Maxwell model. The Giesekus model in not TSS for timescales shorter than $\tau$ \cite{Holz1999}, which means that equation \ref{eqn:MAOS_G3} does not apply, even though the LVE response tracks the Maxwell model. Nevertheless, analytical expressions for intrinsic MAOS moduli $G_{33}^{*}(\omega)$ have been derived previously \cite{KateGurnon2012,Bharadwaj2015}. With $z = \omega \tau$, $\Tp$ and $\Tpp$ are given by,
\begin{align}
\dfrac{G_{33}^{\prime}(z)}{G} & = \dfrac{\alpha_G  z^4  (-21 + 30z^2 + 51z^4 + 4  \alpha_G  (4 - 17z^2 + 3z^4))}{4(1 + z^2)^3  (1 + 4z^2)  (1 + 9z^2)} \notag \\
\dfrac{G_{33}^{\prime\prime}(z)}{G} & = \dfrac{\alpha_G  z^3  (-3  + 48z^2 + 33z^4 -18z^6 + \alpha_G  (2 - 48z^2 + 46z^4))}{4(1 + z^2)^3  (1 + 4z^2)  (1 + 9z^2)}
\label{eqn:maos_giesekus}
\end{align}
Note that the asymptotic dependence of $G_{33}^{\prime}(\omega) \sim \omega^{-2}$ at large $\omega$ differs from that of $K^{\prime}(\omega) \sim \omega^{0}$ for the Maxwell model, while the other asymptotic dependencies at both small and large $\omega$ are identical.

\begin{figure}
\begin{center}
\includegraphics[scale=0.6]{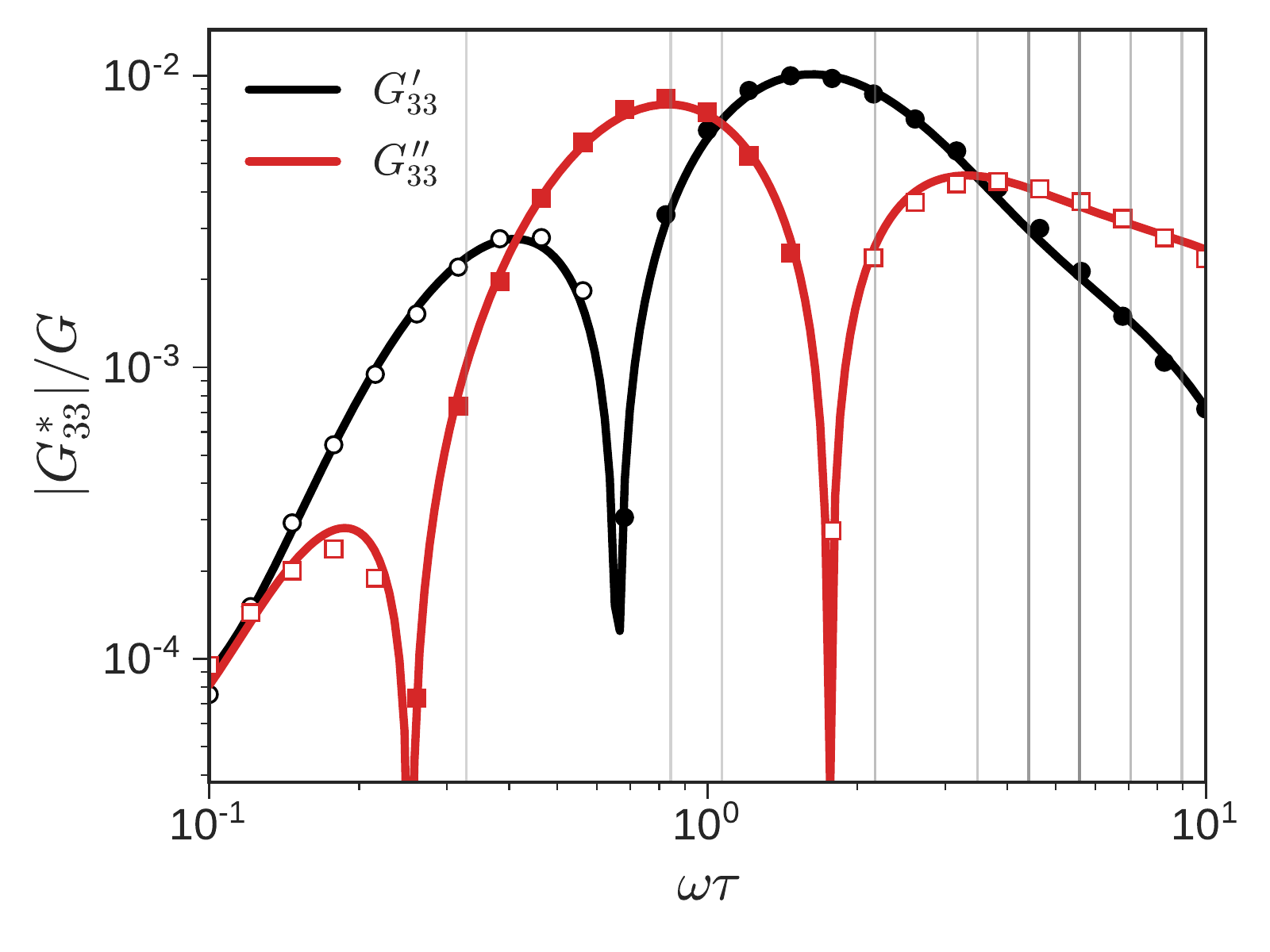}
\caption{Synthetic MAOS moduli $\Tp$ (circles) and $\Tpp$ (squares) generated using the Giesekus model with nonlinear parameter $\alpha_G = 0.2$. Filled (unfilled) symbols denote positive (negative) values. Solid lines of the corresponding color show fits using LASSO regression with $N = 40$ modes. The locations of the nonzero modes $\tau_j$ are indicated by vertical gray lines at frequencies corresponding to $2 \pi/\tau_j$; the darkness of a line increases with mode strength $|g_j|$.  \label{fig:giesekusG3}}
\end{center}
\end{figure}

We generated synthetic experimental data for $G_{33}^*$ from the expressions in equation \ref{eqn:maos_giesekus} using $\alpha_G = 0.2$. Without loss of generality, we set $\tau = 1$ s, and $G = 1$ Pa in our numerical calculations, so that they set the time and modulus scales, respectively. We used $n_d = 25$ logarithmically equispaced  points between $\omega_{\min} \tau = 10^{-1}$ and $\omega_{\max} \tau = 10^{1}$. These are shown by symbols in figure \ref{fig:giesekusG3}. Note that $\Tp$ and $\Tpp$ take both positive and negative values, which are depicted using filled and unfilled symbols, respectively.

To apply the SMEL test, we used the default mode density of $\rho_N = 10$ modes/decade which leads to $N = 40$ logarithmically equispaced $\tau_j$ between $\tau_{\min}/\tau = 10^{-2} $ and $\tau_{\max}/\tau = 10^{2}$. The condition number of the unregularized least-squares feature matrix $\hat{\bm{K}}$ in equation \ref{eqn:chi2_unreg} is $\mathcal{O}(10^{14})$. LASSO adds regularization and makes the problem amenable to numerical solution. The optimal value of the regularization parameter $\alpha$ is found to be $3.4 \times 10^{-4}$. The regression identifies 18 nonzero modes (out of $N = 40$); the modes that lie within the experimental frequency window are indicated in figure \ref{fig:giesekusG3} by vertical gray lines. The darker the line, the larger the magnitude of the corresponding $|g_j|$. The spacing between these lines gives a visual sense of the mode density. Solid lines are fits using these nonzero modes in equation \ref{eqn:drs3}. The agreement between the fitted curves and data is excellent as evidenced by a coefficient of determination value of $R^2 = 0.997$.

It is worthwhile to pause and highlight the advantages of LASSO regression. Recall that two vexing questions that complicate the extraction of the linear relaxation spectrum (DRS) from $\Gst$ are how to select a parsimonious $N$, and where to place the modes $\tau_j$? If $\tau_j$ are not pre-specified, nonlinear least-squares regression, which is computationally costly, has to be performed. LASSO regression allows us to specify a large number of candidate modes $\tau_j$; it completely frees us from the two questions that complicate the calculation of the DRS. At sufficiently high mode density, the modes are closely spaced, which ensures that the relevant timescales are included in the set $\{\tau_j\}$. The regression is robust and automatically discards redundant modes. In the example shown in figure \ref{fig:giesekusG3}, only 18 or 45\% of the originally specified $N = 40$ modes were retained.

\begin{figure}
\begin{center}
\includegraphics[scale=0.6]{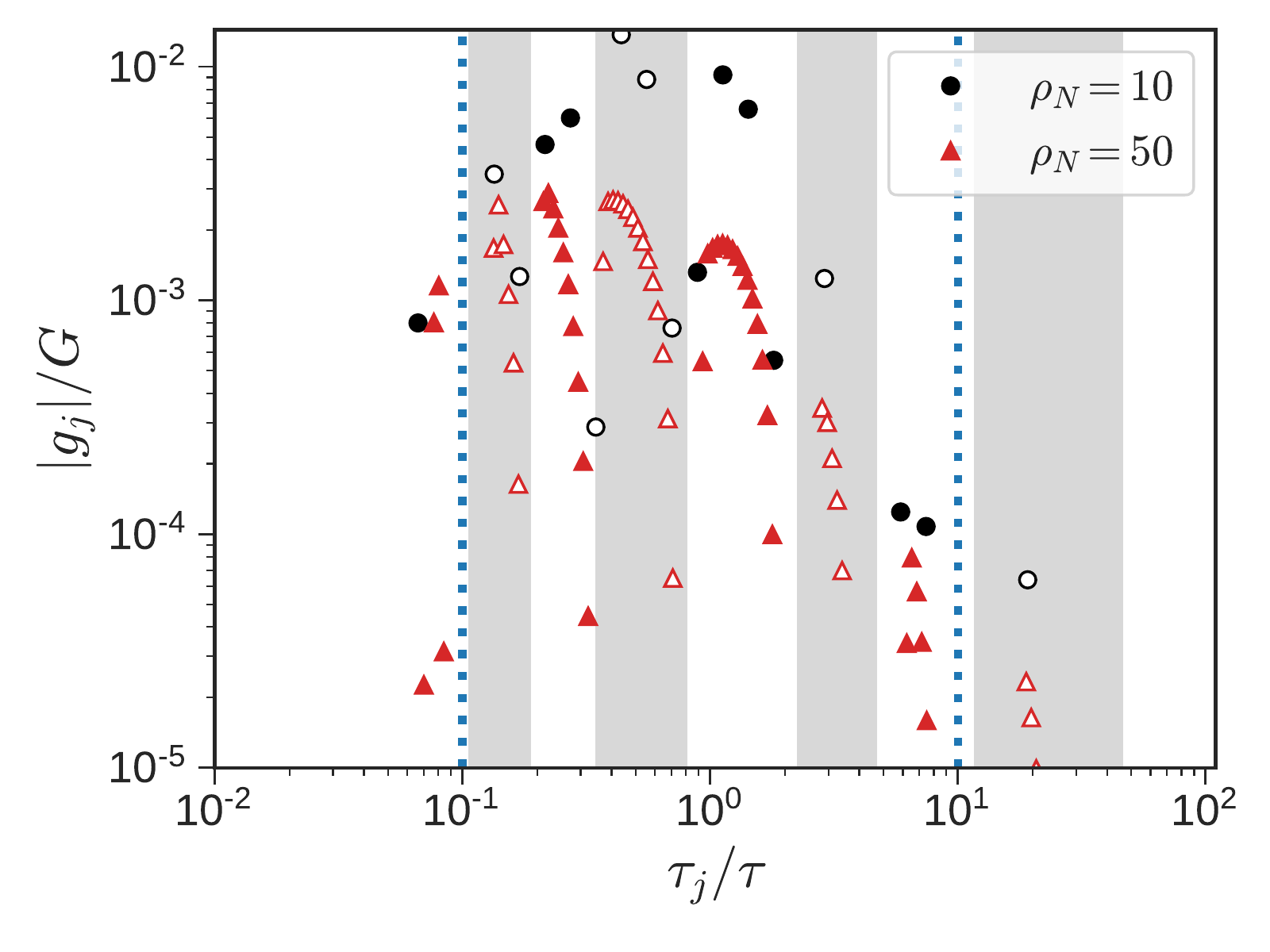}
\caption{The nonzero modes used to fit data in figure \ref{fig:giesekusG3} with mode density of $\rho_N = 10$ modes/decade ($N = 40$) are indicated by black circles. Filled (unfilled) symbols denote positive (negative) values. Shaded patches roughly identify regions where $g_j$ is negative. When the mode density is increased to $\rho_N = 50$ ($N = 200$), we obtain  results depicted by the red triangles. Dotted vertical lines mark the timescales corresponding to the data, viz. $(\omega_{\max} \tau)^{-1}$ and $(\omega_{\min} \tau)^{-1}$. \label{fig:giesekusModes}}
\end{center}
\end{figure}

Figure \ref{fig:giesekusModes} depicts the location ($\tau_j$) and strength ($|g_j|$) of the modes for $N = 40$ ($\rho_N$ = 10 modes/decade) obtained from fitting the data in fig. \ref{fig:giesekusG3}. Note that some of the coefficients $g_j$ are negative, and indicated by open symbols. The majority of the modes identified fall within the range of the experimental data demarcated by the dotted vertical lines. Nevertheless, a non-negligible fraction of the modes lie beyond this range; this situation is also observed in fitting DRS to LVE measurements. 

At $N = 40$ the spacing between successive modes $\tau_{j+1}/\tau_{j} \approx 1.3$.
To test the robustness of the SMEL test to large $N$, we run a numerical experiment by increasing the number of modes to $N = 200$ ($\rho_N = 50$). All but 69 (35\%) of these modes, shown by triangles in figure \ref{fig:giesekusModes}, are discarded as unimportant. The consistency between the locations of $\tau_j$, the sign and relative magnitudes of $g_j$, and the relative independence regularization parameter $\alpha$ with $N$ is reassuring. Note that this large value of $\rho_N =$ 50 modes/decade, which leads to $\tau_{j+1}/\tau_{j} \approx 1.05$, is practically close to the continuous  limit and probably excessive. However, it is shown here to highlight one of the strengths of LASSO regression: its ability to gracefully cope with a large number of modes or degrees of freedom.

\begin{figure}
\begin{center}
\includegraphics[scale=0.6]{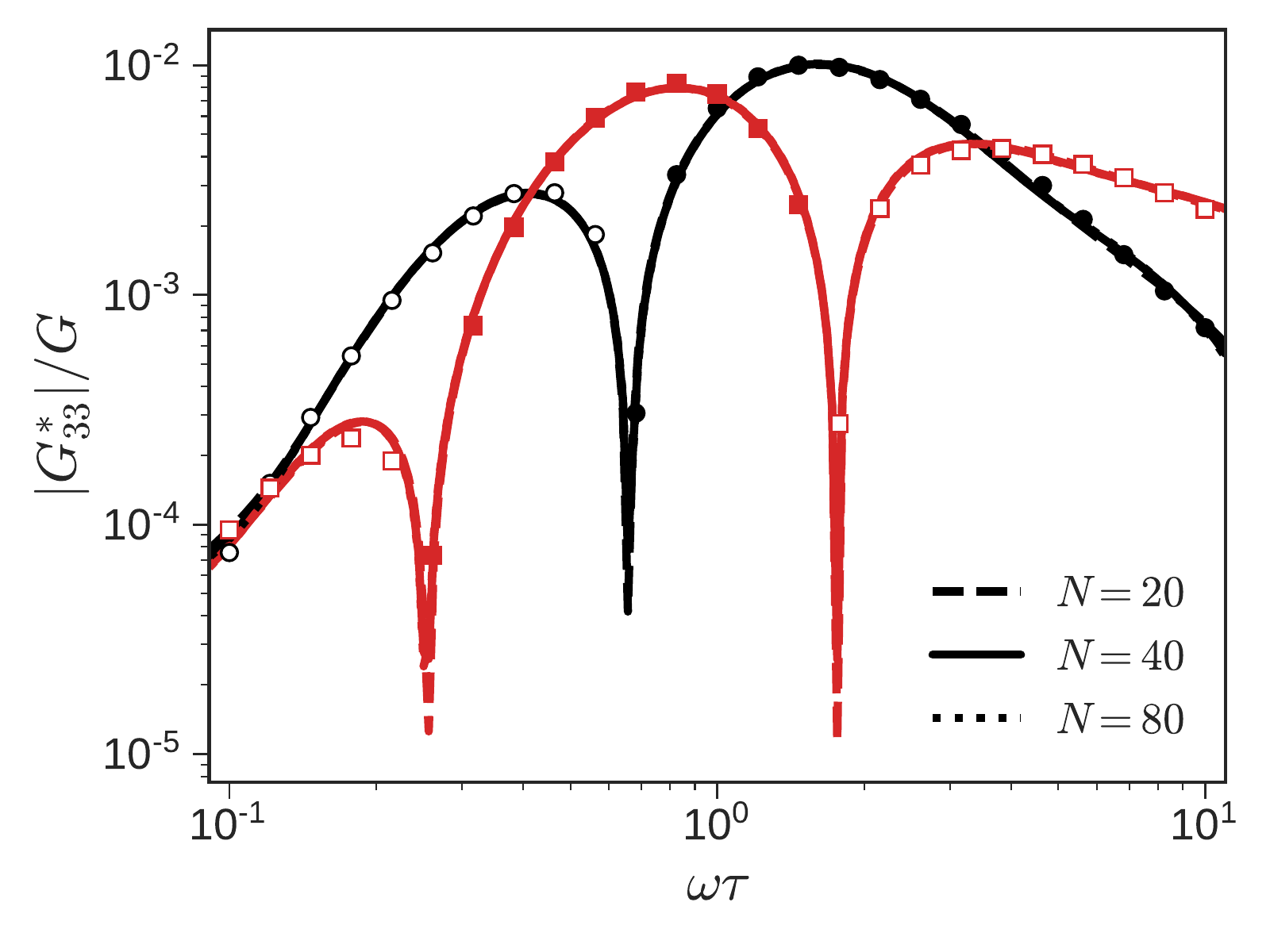}
\caption{Comparison of fits using different choices for the number of initial modes for the Giesekus model shown in fig. \ref{fig:giesekusG3}. The three predictions nearly overlap with each other, and are hard to distinguish.  \label{fig:giesekusCompare}}
\end{center}
\end{figure}

Despite these similarities there is one key difference: computational cost. The $N = 40$ calculation took about 0.17s, whereas the $N = 200$ calculation took 1.45s on a desktop computer with an Intel i7-6700 (3.40GHz) CPU. This trend is expected since the cost of the underlying linear least squares problem asymptotically scales as $\mathcal{O}(N^3)$. The ability of SMEL test appears to be insensitive to $\rho_N$ or $N$. For example using $N = 20$ produces a fit that is visually indistinguishable (see figure \ref{fig:giesekusCompare}) from the $N = 40$ fit reported in figure \ref{fig:giesekusG3}, or larger values of $N$.

These findings may be summarized as follows: the SMEL test (i) correctly identifies KKR compliance of $\Tst$ data even when it is not TSS, (ii) it is efficient; the computational cost is typically $\mathcal{O}(0.1s)$, (iii) it works when experimental data is available on a finite frequency window, and (iv) it is robust and practically insensitive to large $N$; however the asymptotic computational cost increases roughly as $\mathcal{O}(N^3)$.

\subsection{Time-Strain Separable Power-Law Material}

Multiscale complex fluids such as polydisperse and/or branched polymer melts and solutions \cite{Larson1985}, structured food materials \cite{Campanella1987, Poon2016}, the critical gel state in polymeric or colloidal gels \cite{Rathinaraj2021, Suman2021, Suman2020, Winter1997}, etc., show power-law dependence of relaxation modulus, $G(t)=S{t}^{-n}$, over a certain range of timescales $t$. Here, $n \in (0, 1)$ is the power-law exponent, and the quasi-property $S$ has units of Pa$\cdot$s$^{n}$ and characterizes material stiffness. The corresponding storage and loss moduli are \cite{Winter1997},
\begin{align}
\Gp & = \dfrac{\pi S}{2 \Gamma(n)} \dfrac{\omega^n}{\sin(n \pi/2)},\notag\\
\Gpp & = \dfrac{\pi S}{2 \Gamma(n)} \dfrac{\omega^n}{\cos(n \pi/2)}.
\label{eqn:powerlawLVE}
\end{align}
 Several such materials are known to obey TSS \cite{Larson1985,Bavand2017, Suman2019}. Consequently, their $\Tp$ and $\Tpp$ can be obtained from the LVE (equation \ref{eqn:powerlawLVE}), and the damping function parameter $a$ via equation \ref{eqn:MAOS_G3}. Note that while $\Gst$ obeys linear KKR, the $\Tst$ obtained from the LVE assuming TSS violates the MAOS KKR. The weak dependence of the $\Tp$ and $\Tpp$ at low frequencies ($\omega^{n}$) results in a non-integrable singularity in equation \ref{eqn:kk_maos} at $u = 0$. In practice, this issue is moot because power-law behavior is confined to a finite domain of frequencies \cite{Larson1985, Rathinaraj2021}.  

\begin{figure}
\begin{center}
\includegraphics[scale=0.6]{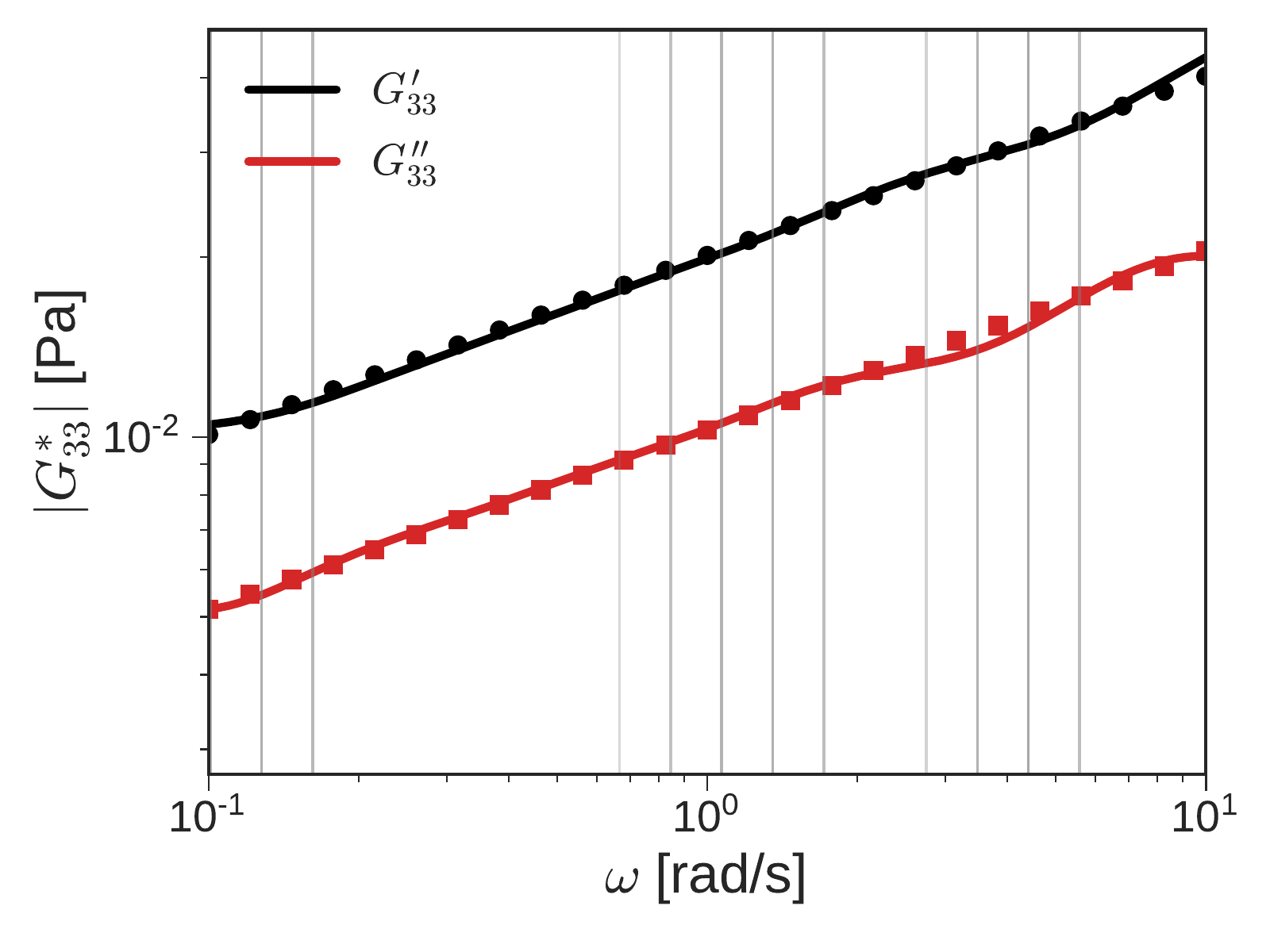}
\caption{Synthetic MAOS moduli $\Tp$ (circles) and $\Tpp$ (squares) generated using the TSS power-law model (equations \ref{eqn:MAOS_G3} and \ref{eqn:powerlawLVE}) with parameters $n=0.3$, $S$ = 1 Pa$\cdot$s$^{n}$, and $a = -0.1$. 5\% noise is added to the pristine data. Solid lines of the corresponding color show fits with $N = 40$ modes. The locations and strengths of the nonzero modes are indicated by vertical gray lines as in figure \ref{fig:giesekusG3}. \label{fig:PLnoise}} 
\end{center}
\end{figure}

Nevertheless, the question of assessing the KKR compliance of power-law behavior experimentally observed over a finite frequency window is both relevant and important. Here, we generate synthetic data between $\omega_{\min} = 10^{-1}$ rad/s and $\omega_{\max} = 10^{1}$ rad/s with $n_d = 25$ data points. We assume $n = 0.3$, $S = 1.0$ Pa$\cdot$s$^n$, and damping function parameter $a = -0.1$. In figure \ref{fig:PLnoise}, we contaminate this ``pristine'' $\Tp$ and $\Tpp$ with 5\% random noise. This noisy dataset is generated by multiplying the pristine data with independent, normally distributed random numbers with mean equal to one, and standard deviation equal to 0.05.

As before, we use $N = 40$ modes and obtain the fit shown by the solid lines in figure \ref{fig:PLnoise}. The agreement between the experimental and fitted data is quite reasonable, and sports a $R^2$ of 0.97. This example demonstrates the robustness of SMEL test to noisy data. Even though the regressed curves have some wiggles, the goal of data validation is accomplished. Note that if the synthetic data is not contaminated by noise, the quality of the fit improves. Thus, this example also shows that over a finite range, power-law behavior can be well-described by a sum of MAOS Maxwell elements.

\begin{figure}
\begin{center}
\includegraphics[scale=0.6]{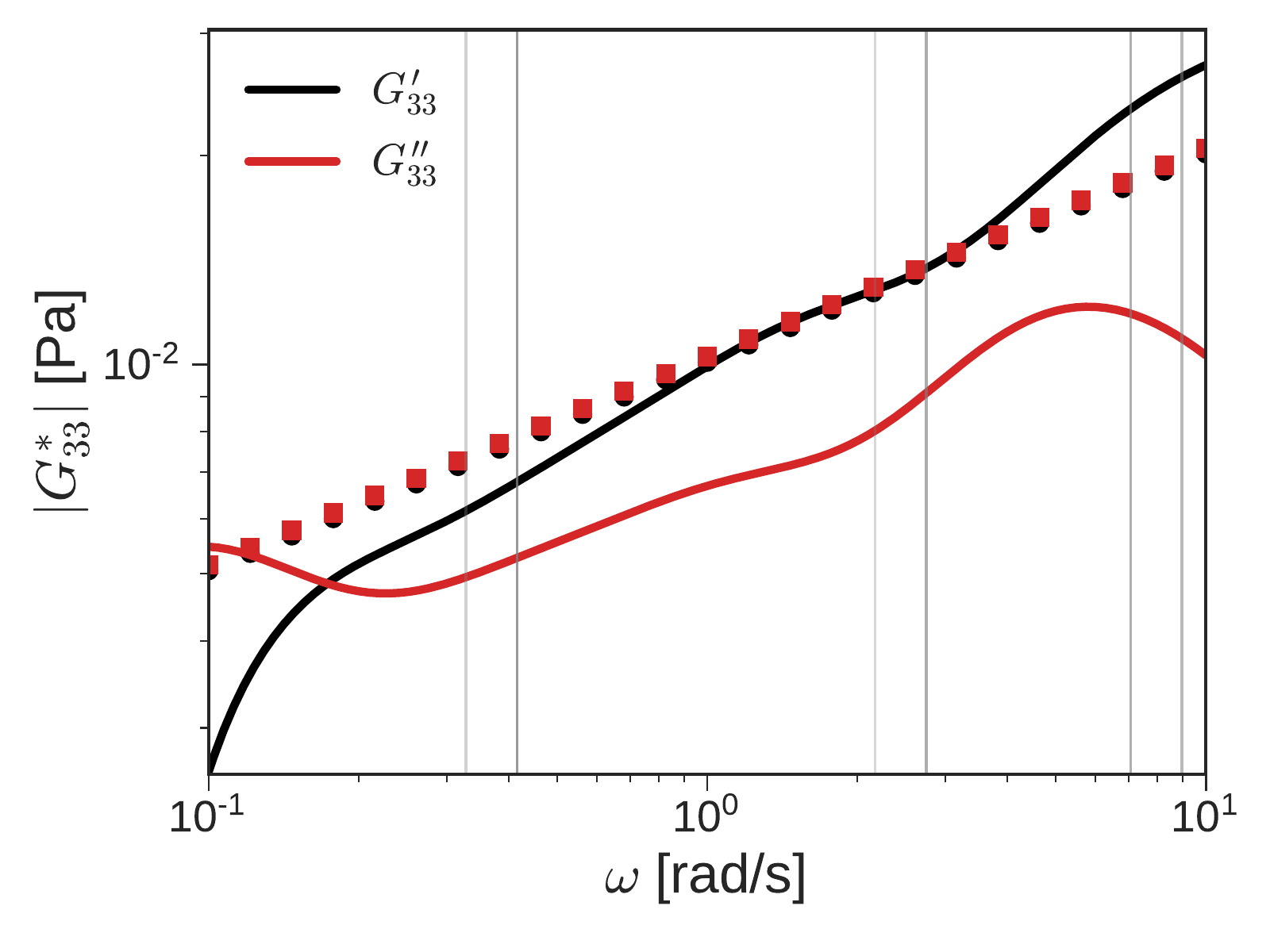}
\caption{Synthetic MAOS moduli $\Tp$ (circles) and $\Tpp$ (squares) generated using the TSS power-law model with parameters $n=0.3$, $S$ = 1 Pa$\cdot$s$^{n}$, and $a = -0.1$. $\Tp$ is then artificially shifted downwards by a factor of two. Solid lines of the corresponding color show fits with $N = 40$ modes. The locations and strengths of the nonzero modes are indicated by vertical gray lines as in figure \ref{fig:giesekusG3}.  \label{fig:PLshift}}
\end{center}
\end{figure}

Up to this point, all synthetic data were generated from analytical expressions for $\Tst$. Therefore, the $\Tp$ and $\Tpp$ were consistent with KKR by default. What we have shown thus far then is that the SMEL test correctly identifies datasets that obey KKR. To test its performance on data that violate KKR, we generate an ``invalid'' dataset using the same parameters as used above for fig. \ref{fig:PLnoise}, except for the noise (including or excluding noise does not change results). We artificially shift the $\Tp$ curve downwards by a factor of two as shown in figure \ref{fig:PLshift}. The $\Tpp$ curve is left untouched.

We use the SMEL test to analyze this data.  As shown in figure \ref{fig:PLshift}, the fits do not agree with the shifted experimental data. Here we used $N = 40$, but increasing $N$ does not improve the agreement as might be expected from figure \ref{fig:giesekusCompare}, which demonstrates that the quality of the fit is  insensitive to $N$. Furthermore, $R^2 = -0.87$ is below any reasonable threshold.  It provides a quantiative proxy for what is visually obvious, leading us to declare that the data is not KKR compliant.

These findings may be summarized as follows: the SMEL test (i) can model power-law behavior over finite frequency windows using Maxwell elements, (ii) it is robust to noise in the data, and (iii) it correctly identifies datasets that are consistent and inconsistent with KKR. When the level of noise is not too large, numerical experiments conducted thus far demonstrate that the SMEL test does not suffer from either false positives or false negatives.

\subsection{Experimental Data}

\begin{figure}
\begin{center}
\includegraphics[scale=0.6]{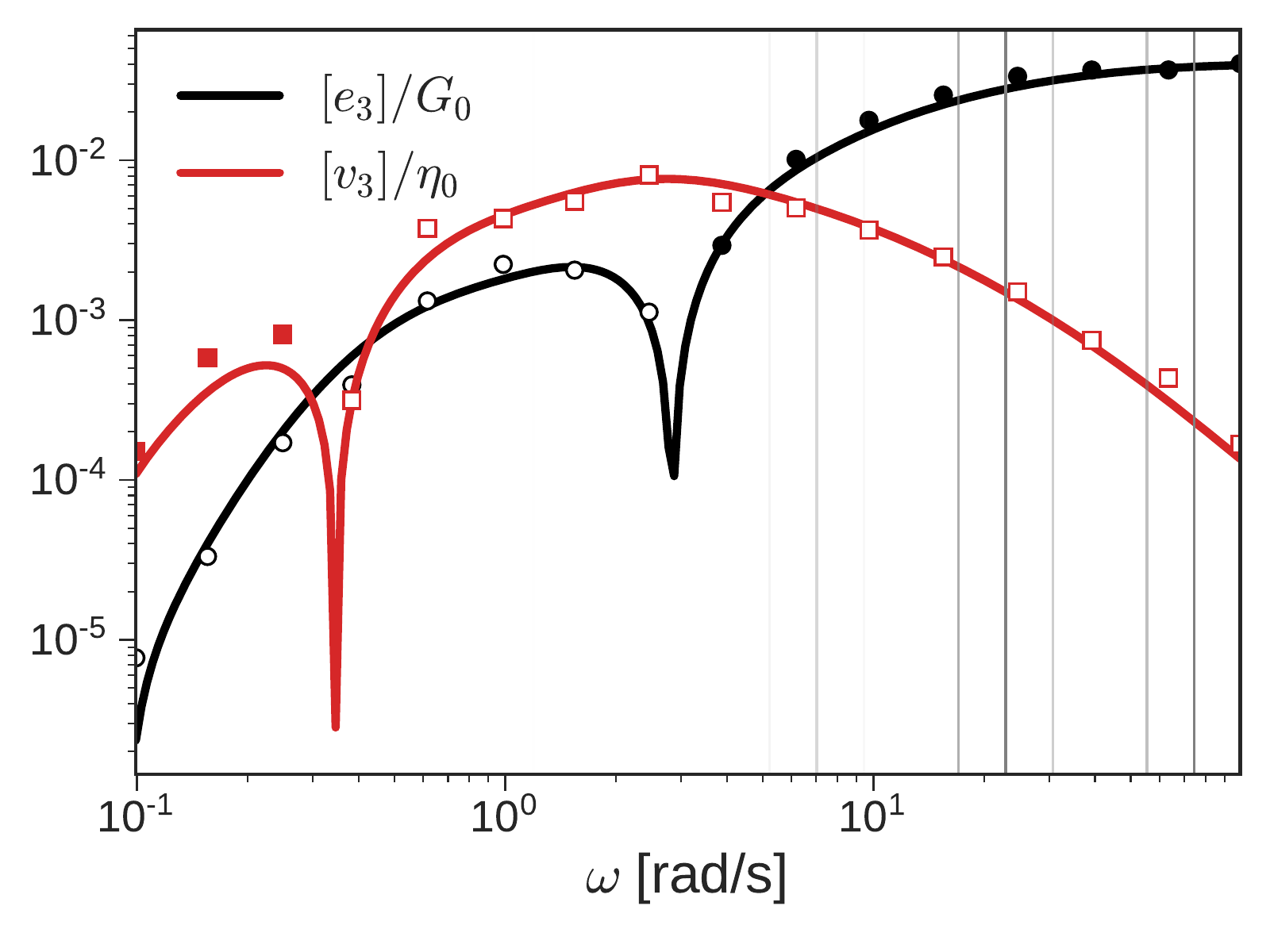}
\caption{Experimental data on a PVA-borax system showing $[e_3] = -\Tp$ (circles) and $[v_3] = \Tpp/\omega$ (squares) \cite{Ewoldt2013, Bharadwaj2015, bharadwaj2016dissertation}. Filled (unfilled) symbols denote positive (negative) values. Solid lines of the corresponding color show fits using $N = 40$. The locations and strengths of the nonzero modes are indicated by vertical gray lines as in figure \ref{fig:giesekusG3}.\cite{Shanbhag2021} \label{fig:EB}}
\end{center}
\end{figure}

Now that we have demonstrated that the SMEL test works quite well on synthetically generated data, we move on to analyze real experimental data. The first experimental report of $G_{33}^{*}$ in the literature is due to Davis and Macosko \cite{Davis1978}. 
However, systematic measurements of frequency-dependent MAOS signatures are more recent. Bharadwaj and Ewoldt reported LVE and MAOS moduli of an aqueous solution of 2.75 wt\% poly vinyl-alcohol (PVA) mixed with 1.25 wt\% sodium tetraborate (borax) \cite{Ewoldt2013, Bharadwaj2015, bharadwaj2016dissertation}. Thermoreversible cross-links between the PVA and borax units endow the material with interesting rheological properties. The LVE signature is simple, and can be nearly approximated by a single Maxwell element. However, common constitutive models do not anticipate the sign changes of MAOS moduli for this system \cite{Bharadwaj2015}. A new network model with non-Hookean springs called the strain-stiffening temporary network model was developed to account for these sign changes \cite{Bharadwaj2017}. From LVE measurements the modulus and zero-shear viscosity were estimated to be $G_0 = 1546 \pm 13$ Pa, and $\eta_0 = 745 \pm 12$ Pa$\cdot$s \cite{bharadwaj2016dissertation}.

Figure \ref{fig:EB} shows the experimentally determined intrinsic MAOS properties in terms of the third order Chebyshev coefficients $[e_3](\omega) = -\Tp(\omega)$ and $[v_3](\omega) = \Tpp(\omega)/\omega$. Fits obtained using SMEL test with $N = 50$ modes are shown by solid lines. Only about a quarter of these modes are found to be nonzero. Overall the agreement between the experiments and the fits is good, as reflected by an $R^2 = 0.974$. Since this is above the cutoff threshold, it suggests that the experimentally extracted MAOS data are compliant with KKR.

\subsection{Implications for First-Harmonic MAOS Moduli}

These examples demonstrate how the SMEL test can be used to efficiently validate $\Tst$ data. A by-product of this test is the set of Maxwell modes $\mathcal{M} = \{g_ i, \tau_i\}$ which fit the data in accordance with equations \ref{eqn:MAOS_kernel} and \ref{eqn:drs3}. The form of the MAOS kernels used in the SMEL test (equation \ref{eqn:MAOS_kernel}) was inspired by the MAOS moduli for TSS materials. In oscillatory shear experiments, TSS materials occupy a special place. For example, their LAOS response can be computed with spectral accuracy using only $\Gst$ and the damping function $h(\gamma)$ \cite{Shanbhag2021}. The MAOS moduli, $G_{31}^{*}$ and $G_{33}^{*}$, can be analytically obtained from $\Gst$ via equation \ref{eqn:MAOS_G3}. However, when TSS is violated, this link between the MAOS and SAOS moduli is severed. 

With these ideas in mind, we propose a numerical experiment.  Suppose, we consider kernel functions for $G_{31}^{*}$ that are valid for TSS materials (similar to equation \ref{eqn:MAOS_kernel} for $G_{33}^{*}$) as,
\begin{align}
K_{31}^{\prime}(z) & = (-3) \left( 4 k^{\prime}(z) -  k^{\prime}(2z) \right)\notag\\
K_{31}^{\prime\prime}(z) & = (-3) \left( 2 k^{\prime\prime}(z) -  k^{\prime\prime}(2z) \right).
\label{eqn:G31_kernel}
\end{align}
We can use the Maxwell modes $\mathcal{M}$ obtained during the SMEL test, and the kernel functions above, to compute ``predictions" for the first-harmonic MAOS moduli, $P_{31}^{*} = P_{31}^{\prime} + i P_{31}^{\prime \prime}$,
\begin{align}
P_{31}^{\prime}(\omega) & = \sum_{j=1}^{N} g_j K_{31}^{\prime}(\omega \tau_j) \notag\\
P_{31}^{\prime\prime}(\omega) & = \sum_{j=1}^{N} g_j K_{31}^{\prime \prime}(\omega \tau_j).
\label{eqn:G31test}
\end{align}
We can then evaluate the correspondence between the $G_{31}^{*}$ data, and the predictions $P_{31}^{*}$. For TSS materials, we expect $P_{31}^{*} \approx G_{31}^{*}$. For non-TSS materials, we expect this approximation to fail.

\begin{figure}
\begin{center}
\includegraphics[width=\textwidth]{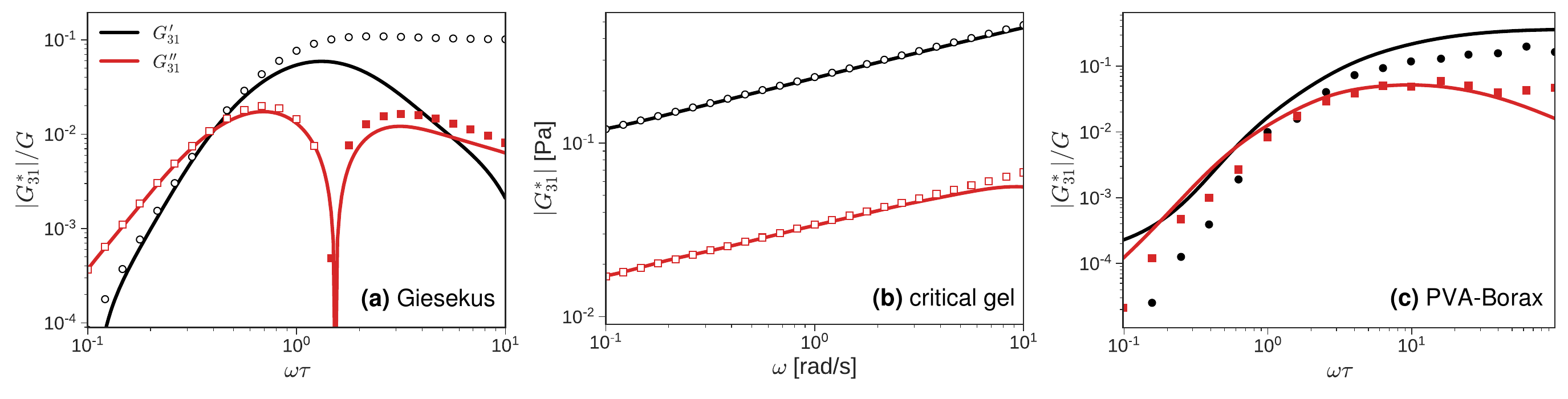}
\caption{Comparisons of synthetic or experimental MAOS moduli $G_{31}^{*}$  (symbols) with predictions $P_{31}^{*}$ obtained from Maxwell modes $\{g_j, \tau_j\}$ for (a) Giesekus model, (b) TSS power-law model, and (c) experimental PVA-Borax system. The Maxwell modes are collected during the SMEL test as part of validating $G_{33}^{*}$ data.  \label{fig:G31}}
\end{center}
\end{figure}

Figure \ref{fig:G31} compares these predictions $P_{31}^{*}$ with the corresponding synthetic or experimental data on $G_{31}^{*}$ for the three different systems considered previously. Different patterns are observed for these three systems, which may be interpreted through the lens of time-strain superposability.

By design, the critical gel in figure \ref{fig:G31}b is TSS. Therefore, it is not surprising that $P_{31}^{*} \approx G_{31}^{*}$ over most of the frequency range. Minor discrepancy is observed near the high-frequency end of the experimental window; this is a manifestation of a familiar phenomenon related to the uncertainty in the extraction of DRS from LVE data \cite{Davies1997}. The disagreement between the inferred and experimental $G_{31}^{*}$ in figure \ref{fig:G31}c suggests that the PVA-Borax system is not TSS. Indeed, the specialized network model used to describe this data is non-TSS \cite{Bharadwaj2017, Martinetti2019}. This brings us to figure \ref{fig:G31}a for the Giesekus model. Interestingly, for $\omega \tau \ll 1$, $P_{31}^{*} \approx G_{31}^{*}$. However, for  $\omega \tau \gg 1$ this correspondence breaks down, especially the prediction for $G_{31}^{\prime}$. This would lead us to correctly conclude that the Giesekus model is not TSS.

Many materials are not \textit{strictly} TSS; instead, they exhibit the property of time-strain separability over a range of timescales in step strain experiments. As an illustrative example, consider polymer solutions and melts where chains relax primarily by reptation. However, when chains are rapidly stretched in strong flows, a relaxation mechanism, which operates on a much quicker timescales called chain retraction also gets activated. This phenomenon leads to non-TSS behavior at short timescales. Interestingly, the Giesekus model qualitatively captures this phenomenology. The nonlinear stress relaxation modulus is given by \cite{Holz1999},
\begin{equation}
G_\text{Giesekus}(t, \gamma) = \dfrac{G_0}{e^{t/\tau} + 2 \alpha^2 \gamma^2 [1 - \text{cosh}(t/\tau)] + \alpha_G \gamma^2 [e^{t/\tau} - 1]},
\label{eqn:Gtgamma}
\end{equation}
where the characteristic time $\tau$ may be loosely thought of as the reptation time. For $t > \tau$, the contribution of $\gamma$ becomes negligible, and $G(t, \gamma)$ becomes proportional to the LVE response $G(t) = G_0 e^{-t/\tau}$, and obeys TSS. In this regime, the damping function is given by $h_\text{Giesekus}(\gamma) = (	\alpha_G(1-\alpha_G)\gamma^2)^{-1}$. However, for $t \lesssim \tau$, TSS is violated. The partial agreement of $P_{31}^{*}$ and $G_{31}^{*}$ in figure \ref{fig:G31}a is a direct reflection of this fact. At low-frequencies $\omega \tau \ll 1$, corresponding to long timescales in $G(t, \gamma)$, the Giesekus model obeys TSS.

Thus, as a by-product the SMEL test can also be used to explore the question of time-strain separability. From figure \ref{fig:G31}, we argue that it can identify TSS and non-TSS materials. Perhaps, more importantly, it has the potential to identify the range of timescales over which some materials are TSS. Note that this inference can also be drawn directly from LVE moduli using equation \ref{eqn:MAOS_G3}. However, direct application of these formulae requires knowledge of the damping function at small strains (the parameter $a$), which involves performing multiple step-strain experiments. The SMEL test method avoids this additional work.

\section{Summary and Conclusions}

The third-harmonic MAOS modulus $G_{33}^{*}$ is extensively used to glean insights into materials that are not immediately visible in LVE data. However, measurement of $G_{33}^{*}$ in experiments is tedious, and fraught with several potential sources of error. Thus, it is important to validate the experimental data, before it can be interpreted.

With this motivation, we proposed a new method called the SMEL test to assess the compliance of $G_{33}^{*}$ with nonlinear KKR. It is inspired by the approach employed to check the consistency of LVE data with linear KKR using Maxwell elements.  In the SMEL test, $G_{33}^{*}$ is expressed as a sum of a large number (approximately 10 modes/decade of frequency) of MAOS kernels inspired by TSS Maxwell elements. It converts the problem of data validation to a linear least squares problem. The ill-conditioning of this problem is fixed using a statistical technique called LASSO, which appends an $L_1$ regularization term to the objective function. LASSO automatically selects a parsimonious set of modes.

The SMEL test is applied to the MAOS response of the Giesekus model, which is not TSS, a TSS power-law model, and an experimental system containing cross-linked polymers, which exhibits a non-standard MAOS fingerprint. The SMEL test work successfully across this broad range of materials and models. It successfully copes with noisy data, and can correctly identify datasets that violate nonlinear KKR. Despite its power and versatility, the SMEL test is simple, barely requiring 20 lines of code, and efficient, requiring runtimes of only a fraction of a second in most cases. Furthermore, the time-strain separability of the material under investigation can be quantified as a byproduct, without running additional step-strain experiments to measure the damping function.

\section*{Supplementary Material}

See supplementary material for the Python code used to run the SMEL Test.

\section*{Acknowledgments}

This work is based in part upon work supported by the National Science Foundation under grant no. NSF DMR-1727870 (SS). YMJ acknowledges financial support from Science and Engineering Research Board (SERB), Department of Science and Technology, Government of India. 

\section*{Data Availability Statement}
The Python code for the SMEL test is listed in supplementary material. Other data that support the findings of this study are available from the corresponding author upon reasonable request.

\bibliography{extracted2}

\begin{thebibliography}{80}%
\makeatletter
\providecommand \@ifxundefined [1]{%
 \@ifx{#1\undefined}
}%
\providecommand \@ifnum [1]{%
 \ifnum #1\expandafter \@firstoftwo
 \else \expandafter \@secondoftwo
 \fi
}%
\providecommand \@ifx [1]{%
 \ifx #1\expandafter \@firstoftwo
 \else \expandafter \@secondoftwo
 \fi
}%
\providecommand \natexlab [1]{#1}%
\providecommand \enquote  [1]{``#1''}%
\providecommand \bibnamefont  [1]{#1}%
\providecommand \bibfnamefont [1]{#1}%
\providecommand \citenamefont [1]{#1}%
\providecommand \href@noop [0]{\@secondoftwo}%
\providecommand \href [0]{\begingroup \@sanitize@url \@href}%
\providecommand \@href[1]{\@@startlink{#1}\@@href}%
\providecommand \@@href[1]{\endgroup#1\@@endlink}%
\providecommand \@sanitize@url [0]{\catcode `\\12\catcode `\$12\catcode
  `\&12\catcode `\#12\catcode `\^12\catcode `\_12\catcode `\%12\relax}%
\providecommand \@@startlink[1]{}%
\providecommand \@@endlink[0]{}%
\providecommand \url  [0]{\begingroup\@sanitize@url \@url }%
\providecommand \@url [1]{\endgroup\@href {#1}{\urlprefix }}%
\providecommand \urlprefix  [0]{URL }%
\providecommand \Eprint [0]{\href }%
\providecommand \doibase [0]{https://doi.org/}%
\providecommand \selectlanguage [0]{\@gobble}%
\providecommand \bibinfo  [0]{\@secondoftwo}%
\providecommand \bibfield  [0]{\@secondoftwo}%
\providecommand \translation [1]{[#1]}%
\providecommand \BibitemOpen [0]{}%
\providecommand \bibitemStop [0]{}%
\providecommand \bibitemNoStop [0]{.\EOS\space}%
\providecommand \EOS [0]{\spacefactor3000\relax}%
\providecommand \BibitemShut  [1]{\csname bibitem#1\endcsname}%
\let\auto@bib@innerbib\@empty
\bibitem [{\citenamefont {Tschoegl}(1989)}]{TschoeglPhenomenological}%
  \BibitemOpen
  \bibfield  {author} {\bibinfo {author} {\bibfnamefont {N.~W.}\ \bibnamefont
  {Tschoegl}},\ }\href@noop {} {\emph {\bibinfo {title} {The phenomenological
  theory of linear viscoelastic behavior: {An} introduction}}},\ \bibinfo
  {edition} {$1^\text{st}$}\ ed.\ (\bibinfo  {publisher} {Springer-Verlag},\
  \bibinfo {address} {Munich, Germany},\ \bibinfo {year} {1989})\BibitemShut
  {NoStop}%
\bibitem [{\citenamefont {Ferry}(1980)}]{Ferry1980}%
  \BibitemOpen
  \bibfield  {author} {\bibinfo {author} {\bibfnamefont {J.~D.}\ \bibnamefont
  {Ferry}},\ }\href@noop {} {\emph {\bibinfo {title} {Viscoelastic properties
  of polymers}}},\ \bibinfo {edition} {$3^\text{rd}$}\ ed.\ (\bibinfo
  {publisher} {John Wiley \& Sons},\ \bibinfo {address} {New York, NY},\
  \bibinfo {year} {1980})\BibitemShut {NoStop}%
\bibitem [{\citenamefont {Cho}(2016)}]{Cho2016}%
  \BibitemOpen
  \bibfield  {author} {\bibinfo {author} {\bibfnamefont {K.~S.}\ \bibnamefont
  {Cho}},\ }\href@noop {} {\emph {\bibinfo {title} {Viscoelasticity of
  Polymers: {T}heory and Numerical Algorithms}}}\ (\bibinfo  {publisher}
  {Springer},\ \bibinfo {address} {Dordrecht, the Netherlands},\ \bibinfo
  {year} {2016})\BibitemShut {NoStop}%
\bibitem [{\citenamefont {de~L.~Kronig}(1926)}]{L.Kronig1926}%
  \BibitemOpen
  \bibfield  {author} {\bibinfo {author} {\bibfnamefont {R.}~\bibnamefont
  {de~L.~Kronig}},\ }\bibfield  {title} {\bibinfo {title} {On the theory of
  dispersion of {X}-rays},\ }\href {https://doi.org/10.1364/JOSA.12.000547}
  {\bibfield  {journal} {\bibinfo  {journal} {J. Opt. Soc. Am.}\ }\textbf
  {\bibinfo {volume} {12}},\ \bibinfo {pages} {547} (\bibinfo {year}
  {1926})}\BibitemShut {NoStop}%
\bibitem [{\citenamefont {Kramers}(1927)}]{Kramers1927}%
  \BibitemOpen
  \bibfield  {author} {\bibinfo {author} {\bibfnamefont {H.~A.}\ \bibnamefont
  {Kramers}},\ }\bibfield  {title} {\bibinfo {title} {La diffusion de la
  lumiere par les atomes},\ }in\ \href@noop {} {\emph {\bibinfo {booktitle}
  {Atti Cong. Intern. Fisica, Como}}},\ Vol.~\bibinfo {volume} {2}\ (\bibinfo
  {year} {1927})\ pp.\ \bibinfo {pages} {545--557}\BibitemShut {NoStop}%
\bibitem [{\citenamefont {Shanbhag}\ and\ \citenamefont {Joshi}(2022)}]{kkr1}%
  \BibitemOpen
  \bibfield  {author} {\bibinfo {author} {\bibfnamefont {S.}~\bibnamefont
  {Shanbhag}}\ and\ \bibinfo {author} {\bibfnamefont {Y.~M.}\ \bibnamefont
  {Joshi}},\ }\href@noop {} {\bibinfo {title} {{Kramers-Kronig} relations for
  nonlinear rheology: 1. {General Expression} and implications}} (\bibinfo
  {year} {2022})\BibitemShut {NoStop}%
\bibitem [{\citenamefont {Peiponen}\ and\ \citenamefont
  {Vartiainen}(1991)}]{Peiponen1991}%
  \BibitemOpen
  \bibfield  {author} {\bibinfo {author} {\bibfnamefont {K.-E.}\ \bibnamefont
  {Peiponen}}\ and\ \bibinfo {author} {\bibfnamefont {E.~M.}\ \bibnamefont
  {Vartiainen}},\ }\bibfield  {title} {\bibinfo {title} {{Kramers-Kronig}
  relations in optical data inversion},\ }\href
  {https://doi.org/10.1103/PhysRevB.44.8301} {\bibfield  {journal} {\bibinfo
  {journal} {Phys. Rev. B}\ }\textbf {\bibinfo {volume} {44}},\ \bibinfo
  {pages} {8301} (\bibinfo {year} {1991})}\BibitemShut {NoStop}%
\bibitem [{\citenamefont {Lucarini}\ \emph {et~al.}(2005)\citenamefont
  {Lucarini}, \citenamefont {Saarinen}, \citenamefont {Peiponen},\ and\
  \citenamefont {Vartiainen}}]{Lucarini2005}%
  \BibitemOpen
  \bibfield  {author} {\bibinfo {author} {\bibfnamefont {V.}~\bibnamefont
  {Lucarini}}, \bibinfo {author} {\bibfnamefont {J.~J.}\ \bibnamefont
  {Saarinen}}, \bibinfo {author} {\bibfnamefont {K.-E.}\ \bibnamefont
  {Peiponen}},\ and\ \bibinfo {author} {\bibfnamefont {E.~M.}\ \bibnamefont
  {Vartiainen}},\ }\href {https://doi.org/https://doi.org/10.1007/b138913}
  {\emph {\bibinfo {title} {Kramers-Kronig relations in optical materials
  research}}},\ \bibinfo {edition} {1st}\ ed.,\ Vol.\ \bibinfo {volume} {110}\
  (\bibinfo  {publisher} {Springer},\ \bibinfo {address} {Berlin, Germany},\
  \bibinfo {year} {2005})\BibitemShut {NoStop}%
\bibitem [{\citenamefont {Gross}(1941)}]{Gross1941}%
  \BibitemOpen
  \bibfield  {author} {\bibinfo {author} {\bibfnamefont {B.}~\bibnamefont
  {Gross}},\ }\bibfield  {title} {\bibinfo {title} {On the theory of dielectric
  loss},\ }\href {https://doi.org/10.1103/PhysRev.59.748} {\bibfield  {journal}
  {\bibinfo  {journal} {Phys. Rev.}\ }\textbf {\bibinfo {volume} {59}},\
  \bibinfo {pages} {748} (\bibinfo {year} {1941})}\BibitemShut {NoStop}%
\bibitem [{\citenamefont {Boukamp}(2004)}]{Boukamp2004}%
  \BibitemOpen
  \bibfield  {author} {\bibinfo {author} {\bibfnamefont {B.~A.}\ \bibnamefont
  {Boukamp}},\ }\bibfield  {title} {\bibinfo {title} {Electrochemical impedance
  spectroscopy in solid state ionics: {Recent} advances},\ }\href
  {https://doi.org/https://doi.org/10.1016/j.ssi.2003.07.002} {\bibfield
  {journal} {\bibinfo  {journal} {Solid State Ionics}\ }\textbf {\bibinfo
  {volume} {169}},\ \bibinfo {pages} {65} (\bibinfo {year} {2004})},\ \bibinfo
  {note} {proceedings of the Annual Meeting of International Society of
  Electrochemistry}\BibitemShut {NoStop}%
\bibitem [{\citenamefont {Bode}(1945)}]{Bode1945}%
  \BibitemOpen
  \bibfield  {author} {\bibinfo {author} {\bibfnamefont {H.}~\bibnamefont
  {Bode}},\ }\href {https://books.google.com/books?id=fDv0tQEACAAJ} {\emph
  {\bibinfo {title} {Network analysis and feedback amplifier design}}}\
  (\bibinfo  {publisher} {D. Van Nostrand Company},\ \bibinfo {address}
  {Princeton, New Jersey},\ \bibinfo {year} {1945})\BibitemShut {NoStop}%
\bibitem [{\citenamefont {Silva}\ and\ \citenamefont
  {Gross}(1941)}]{Silva1941}%
  \BibitemOpen
  \bibfield  {author} {\bibinfo {author} {\bibfnamefont {H.}~\bibnamefont
  {Silva}}\ and\ \bibinfo {author} {\bibfnamefont {B.}~\bibnamefont {Gross}},\
  }\bibfield  {title} {\bibinfo {title} {Some measurements on the validity of
  the principle of superposition in solid dielectrics},\ }\href
  {https://doi.org/10.1103/PhysRev.60.684} {\bibfield  {journal} {\bibinfo
  {journal} {Phys. Rev.}\ }\textbf {\bibinfo {volume} {60}},\ \bibinfo {pages}
  {684} (\bibinfo {year} {1941})}\BibitemShut {NoStop}%
\bibitem [{\citenamefont {Lovell}(1974)}]{Lovell1974}%
  \BibitemOpen
  \bibfield  {author} {\bibinfo {author} {\bibfnamefont {R.}~\bibnamefont
  {Lovell}},\ }\bibfield  {title} {\bibinfo {title} {Application of
  {Kramers-Kronig} relations to the interpretation of dielectric data},\ }\href
  {https://doi.org/10.1088/0022-3719/7/23/024} {\bibfield  {journal} {\bibinfo
  {journal} {J. Phys. C: Solid State Phys.}\ }\textbf {\bibinfo {volume} {7}},\
  \bibinfo {pages} {4378} (\bibinfo {year} {1974})}\BibitemShut {NoStop}%
\bibitem [{\citenamefont {Davis}\ and\ \citenamefont
  {Rabinowitz}(1984)}]{Davis1984}%
  \BibitemOpen
  \bibfield  {author} {\bibinfo {author} {\bibfnamefont {P.~J.}\ \bibnamefont
  {Davis}}\ and\ \bibinfo {author} {\bibfnamefont {P.}~\bibnamefont
  {Rabinowitz}},\ }\href
  {https://doi.org/https://doi.org/10.1016/B978-0-12-206360-2.50007-8} {\emph
  {\bibinfo {title} {Methods of Numerical Integration}}},\ \bibinfo {edition}
  {second edition}\ ed.\ (\bibinfo  {publisher} {Academic Press},\ \bibinfo
  {year} {1984})\ pp.\ \bibinfo {pages} {1--50}\BibitemShut {NoStop}%
\bibitem [{\citenamefont {Amari}\ and\ \citenamefont
  {Bornemann}(1995)}]{Amari1995}%
  \BibitemOpen
  \bibfield  {author} {\bibinfo {author} {\bibfnamefont {S.}~\bibnamefont
  {Amari}}\ and\ \bibinfo {author} {\bibfnamefont {J.}~\bibnamefont
  {Bornemann}},\ }\bibfield  {title} {\bibinfo {title} {Efficient numerical
  computation of singular integrals with applications to electromagnetics},\
  }\href {https://doi.org/10.1109/8.475113} {\bibfield  {journal} {\bibinfo
  {journal} {IEEE Trans. Antennas Propag.}\ }\textbf {\bibinfo {volume} {43}},\
  \bibinfo {pages} {1343} (\bibinfo {year} {1995})}\BibitemShut {NoStop}%
\bibitem [{\citenamefont {Urquidi-Macdonald}\ \emph {et~al.}(1986)\citenamefont
  {Urquidi-Macdonald}, \citenamefont {Real},\ and\ \citenamefont
  {Macdonald}}]{Urquidi-Macdonald1986}%
  \BibitemOpen
  \bibfield  {author} {\bibinfo {author} {\bibfnamefont {M.}~\bibnamefont
  {Urquidi-Macdonald}}, \bibinfo {author} {\bibfnamefont {S.}~\bibnamefont
  {Real}},\ and\ \bibinfo {author} {\bibfnamefont {D.~D.}\ \bibnamefont
  {Macdonald}},\ }\bibfield  {title} {\bibinfo {title} {Application of
  {Kramers-Kronig} transforms in the analysis of electrochemical impedance
  data: {II . T}ransformations in the complex plane},\ }\href
  {https://doi.org/10.1149/1.2108332} {\bibfield  {journal} {\bibinfo
  {journal} {J. Electrochem. Soc.}\ }\textbf {\bibinfo {volume} {133}},\
  \bibinfo {pages} {2018} (\bibinfo {year} {1986})}\BibitemShut {NoStop}%
\bibitem [{\citenamefont {Urquidi-Macdonald}\ \emph {et~al.}(1990)\citenamefont
  {Urquidi-Macdonald}, \citenamefont {Real},\ and\ \citenamefont
  {Macdonald}}]{Urquidi-Macdonald1990}%
  \BibitemOpen
  \bibfield  {author} {\bibinfo {author} {\bibfnamefont {M.}~\bibnamefont
  {Urquidi-Macdonald}}, \bibinfo {author} {\bibfnamefont {S.}~\bibnamefont
  {Real}},\ and\ \bibinfo {author} {\bibfnamefont {D.~D.}\ \bibnamefont
  {Macdonald}},\ }\bibfield  {title} {\bibinfo {title} {Applications of
  {Kramers-Kronig} transforms in the analysis of electrochemical impedance data
  - {III. Stability} and linearity},\ }\href
  {https://doi.org/https://doi.org/10.1016/0013-4686(90)80010-L} {\bibfield
  {journal} {\bibinfo  {journal} {Electrochim. Acta}\ }\textbf {\bibinfo
  {volume} {35}},\ \bibinfo {pages} {1559} (\bibinfo {year}
  {1990})}\BibitemShut {NoStop}%
\bibitem [{\citenamefont {King}(2002)}]{King2002}%
  \BibitemOpen
  \bibfield  {author} {\bibinfo {author} {\bibfnamefont {F.~W.}\ \bibnamefont
  {King}},\ }\bibfield  {title} {\bibinfo {title} {Efficient numerical approach
  to the evaluation of {Kramers-Kronig} transforms},\ }\href
  {https://doi.org/10.1364/JOSAB.19.002427} {\bibfield  {journal} {\bibinfo
  {journal} {J. Opt. Soc. Am. B}\ }\textbf {\bibinfo {volume} {19}},\ \bibinfo
  {pages} {2427} (\bibinfo {year} {2002})}\BibitemShut {NoStop}%
\bibitem [{\citenamefont {King}(2007)}]{King2007}%
  \BibitemOpen
  \bibfield  {author} {\bibinfo {author} {\bibfnamefont {F.~W.}\ \bibnamefont
  {King}},\ }\bibfield  {title} {\bibinfo {title} {Numerical evaluation of
  truncated {Kramers-Kronig} transforms},\ }\href
  {https://doi.org/10.1364/JOSAB.24.001589} {\bibfield  {journal} {\bibinfo
  {journal} {J. Opt. Soc. Am. B}\ }\textbf {\bibinfo {volume} {24}},\ \bibinfo
  {pages} {1589} (\bibinfo {year} {2007})}\BibitemShut {NoStop}%
\bibitem [{\citenamefont {Esteban}\ and\ \citenamefont
  {Orazem}(1991)}]{Esteban1991}%
  \BibitemOpen
  \bibfield  {author} {\bibinfo {author} {\bibfnamefont {J.~M.}\ \bibnamefont
  {Esteban}}\ and\ \bibinfo {author} {\bibfnamefont {M.~E.}\ \bibnamefont
  {Orazem}},\ }\bibfield  {title} {\bibinfo {title} {On the application of the
  {Kramers-Kronig} relations to evaluate the consistency of electrochemical
  impedance data},\ }\href {https://doi.org/10.1149/1.2085580} {\bibfield
  {journal} {\bibinfo  {journal} {J Electrochem Soc}\ }\textbf {\bibinfo
  {volume} {138}},\ \bibinfo {pages} {67} (\bibinfo {year} {1991})}\BibitemShut
  {NoStop}%
\bibitem [{\citenamefont {Bakry}\ and\ \citenamefont
  {Klinkenbusch}(2018)}]{Bakry2018}%
  \BibitemOpen
  \bibfield  {author} {\bibinfo {author} {\bibfnamefont {M.}~\bibnamefont
  {Bakry}}\ and\ \bibinfo {author} {\bibfnamefont {L.}~\bibnamefont
  {Klinkenbusch}},\ }\bibfield  {title} {\bibinfo {title} {Using the
  {Kramers-Kronig} transforms to retrieve the conductivity from the effective
  complex permittivity},\ }\href {https://doi.org/10.5194/ars-16-23-2018}
  {\bibfield  {journal} {\bibinfo  {journal} {Adv. Radio Sci.}\ }\textbf
  {\bibinfo {volume} {16}},\ \bibinfo {pages} {23} (\bibinfo {year}
  {2018})}\BibitemShut {NoStop}%
\bibitem [{\citenamefont {Rouleau}\ \emph {et~al.}(2013)\citenamefont
  {Rouleau}, \citenamefont {Deü}, \citenamefont {Legay},\ and\ \citenamefont
  {{Le Lay}}}]{Rouleau2013}%
  \BibitemOpen
  \bibfield  {author} {\bibinfo {author} {\bibfnamefont {L.}~\bibnamefont
  {Rouleau}}, \bibinfo {author} {\bibfnamefont {J.-F.}\ \bibnamefont {Deü}},
  \bibinfo {author} {\bibfnamefont {A.}~\bibnamefont {Legay}},\ and\ \bibinfo
  {author} {\bibfnamefont {F.}~\bibnamefont {{Le Lay}}},\ }\bibfield  {title}
  {\bibinfo {title} {Application of {Kramers-Kronig} relations to
  time-temperature superposition for viscoelastic materials},\ }\href
  {https://doi.org/https://doi.org/10.1016/j.mechmat.2013.06.001} {\bibfield
  {journal} {\bibinfo  {journal} {Mech. Mater.}\ }\textbf {\bibinfo {volume}
  {65}},\ \bibinfo {pages} {66} (\bibinfo {year} {2013})}\BibitemShut {NoStop}%
\bibitem [{\citenamefont {Erwin}\ \emph {et~al.}(2010)\citenamefont {Erwin},
  \citenamefont {Rogers}, \citenamefont {Cloitre},\ and\ \citenamefont
  {Vlassopoulos}}]{Erwin2010}%
  \BibitemOpen
  \bibfield  {author} {\bibinfo {author} {\bibfnamefont {B.~M.}\ \bibnamefont
  {Erwin}}, \bibinfo {author} {\bibfnamefont {S.~A.}\ \bibnamefont {Rogers}},
  \bibinfo {author} {\bibfnamefont {M.}~\bibnamefont {Cloitre}},\ and\ \bibinfo
  {author} {\bibfnamefont {D.}~\bibnamefont {Vlassopoulos}},\ }\bibfield
  {title} {\bibinfo {title} {Examining the validity of strain-rate frequency
  superposition when measuring the linear viscoelastic properties of soft
  materials},\ }\href {https://doi.org/10.1122/1.3301247} {\bibfield  {journal}
  {\bibinfo  {journal} {J. Rheol.}\ }\textbf {\bibinfo {volume} {54}},\
  \bibinfo {pages} {187} (\bibinfo {year} {2010})}\BibitemShut {NoStop}%
\bibitem [{\citenamefont {Wyss}\ \emph {et~al.}(2007)\citenamefont {Wyss},
  \citenamefont {Miyazaki}, \citenamefont {Mattsson}, \citenamefont {Hu},
  \citenamefont {Reichman},\ and\ \citenamefont {Weitz}}]{Wyss2007}%
  \BibitemOpen
  \bibfield  {author} {\bibinfo {author} {\bibfnamefont {H.~M.}\ \bibnamefont
  {Wyss}}, \bibinfo {author} {\bibfnamefont {K.}~\bibnamefont {Miyazaki}},
  \bibinfo {author} {\bibfnamefont {J.}~\bibnamefont {Mattsson}}, \bibinfo
  {author} {\bibfnamefont {Z.}~\bibnamefont {Hu}}, \bibinfo {author}
  {\bibfnamefont {D.~R.}\ \bibnamefont {Reichman}},\ and\ \bibinfo {author}
  {\bibfnamefont {D.~A.}\ \bibnamefont {Weitz}},\ }\bibfield  {title} {\bibinfo
  {title} {Strain-rate frequency superposition: {A} rheological probe of
  structural relaxation in soft materials},\ }\href
  {https://doi.org/10.1103/PhysRevLett.98.238303} {\bibfield  {journal}
  {\bibinfo  {journal} {Phys. Rev. Lett.}\ }\textbf {\bibinfo {volume} {98}},\
  \bibinfo {pages} {238303} (\bibinfo {year} {2007})}\BibitemShut {NoStop}%
\bibitem [{\citenamefont {Winter}(1997)}]{Winter1997}%
  \BibitemOpen
  \bibfield  {author} {\bibinfo {author} {\bibfnamefont {H.}~\bibnamefont
  {Winter}},\ }\bibfield  {title} {\bibinfo {title} {Analysis of dynamic
  mechanical data: inversion into a relaxation time spectrum and consistency
  check},\ }\href
  {https://doi.org/https://doi.org/10.1016/S0377-0257(96)01512-1} {\bibfield
  {journal} {\bibinfo  {journal} {J. Non-Newtonian Fluid Mech.}\ }\textbf
  {\bibinfo {volume} {68}},\ \bibinfo {pages} {225} (\bibinfo {year} {1997})},\
  \bibinfo {note} {papers presented at the Polymer Melt Rheology
  Conference}\BibitemShut {NoStop}%
\bibitem [{\citenamefont {Boukamp}(1995)}]{Boukamp1995}%
  \BibitemOpen
  \bibfield  {author} {\bibinfo {author} {\bibfnamefont {B.~A.}\ \bibnamefont
  {Boukamp}},\ }\bibfield  {title} {\bibinfo {title} {A linear {Kronig-Kramers}
  transform test for immittance data validation},\ }\href
  {https://doi.org/10.1149/1.2044210} {\bibfield  {journal} {\bibinfo
  {journal} {J Electrochem Soc}\ }\textbf {\bibinfo {volume} {142}},\ \bibinfo
  {pages} {1885} (\bibinfo {year} {1995})}\BibitemShut {NoStop}%
\bibitem [{\citenamefont {Agarwal}\ \emph {et~al.}(1992)\citenamefont
  {Agarwal}, \citenamefont {Orazem},\ and\ \citenamefont
  {Garcia-Rubio}}]{Agarwal1992}%
  \BibitemOpen
  \bibfield  {author} {\bibinfo {author} {\bibfnamefont {P.}~\bibnamefont
  {Agarwal}}, \bibinfo {author} {\bibfnamefont {M.~E.}\ \bibnamefont
  {Orazem}},\ and\ \bibinfo {author} {\bibfnamefont {L.~H.}\ \bibnamefont
  {Garcia-Rubio}},\ }\bibfield  {title} {\bibinfo {title} {Measurement models
  for electrochemical impedance spectroscopy: {I . Demonstration} of
  applicability},\ }\href {https://doi.org/10.1149/1.2069522} {\bibfield
  {journal} {\bibinfo  {journal} {J Electrochem Soc}\ }\textbf {\bibinfo
  {volume} {139}},\ \bibinfo {pages} {1917} (\bibinfo {year}
  {1992})}\BibitemShut {NoStop}%
\bibitem [{\citenamefont {Provencher}(1976)}]{provencher76}%
  \BibitemOpen
  \bibfield  {author} {\bibinfo {author} {\bibfnamefont {S.~W.}\ \bibnamefont
  {Provencher}},\ }\bibfield  {title} {\bibinfo {title} {An eigenfunction
  expansion method for the analysis of exponential decay curves},\ }\href
  {https://doi.org/10.1063/1.432601} {\bibfield  {journal} {\bibinfo  {journal}
  {J. Chem. Phys.}\ }\textbf {\bibinfo {volume} {64}},\ \bibinfo {pages} {2772}
  (\bibinfo {year} {1976})}\BibitemShut {NoStop}%
\bibitem [{\citenamefont {Takeh}\ and\ \citenamefont
  {Shanbhag}(2013)}]{Takeh2013}%
  \BibitemOpen
  \bibfield  {author} {\bibinfo {author} {\bibfnamefont {A.}~\bibnamefont
  {Takeh}}\ and\ \bibinfo {author} {\bibfnamefont {S.}~\bibnamefont
  {Shanbhag}},\ }\bibfield  {title} {\bibinfo {title} {A computer program to
  extract the continuous and discrete relaxation spectra from dynamic
  viscoelastic measurements},\ }\href@noop {} {\bibfield  {journal} {\bibinfo
  {journal} {Applied Rheology}\ }\textbf {\bibinfo {volume} {23}},\ \bibinfo
  {pages} {95} (\bibinfo {year} {2013})}\BibitemShut {NoStop}%
\bibitem [{\citenamefont {Shanbhag}(2019)}]{Shanbhag2019respect}%
  \BibitemOpen
  \bibfield  {author} {\bibinfo {author} {\bibfnamefont {S.}~\bibnamefont
  {Shanbhag}},\ }\bibfield  {title} {\bibinfo {title} {{pyReSpect: A} computer
  program to extract discrete and continuous spectra from stress relaxation
  experiments},\ }\href {https://doi.org/10.1002/mats.201900005} {\bibfield
  {journal} {\bibinfo  {journal} {Macromol. Theory Simul.}\ ,\ \bibinfo {pages}
  {1900005}} (\bibinfo {year} {2019})}\BibitemShut {NoStop}%
\bibitem [{\citenamefont {Shanbhag}(2020)}]{Shanbhag2020}%
  \BibitemOpen
  \bibfield  {author} {\bibinfo {author} {\bibfnamefont {S.}~\bibnamefont
  {Shanbhag}},\ }\bibfield  {title} {\bibinfo {title} {Relaxation spectra using
  nonlinear {Tikhonov} regularization with a {Bayesian} criterion},\ }\href
  {https://doi.org/10.1007/s00397-020-01212-w} {\bibfield  {journal} {\bibinfo
  {journal} {Rheol. Acta}\ }\textbf {\bibinfo {volume} {59}},\ \bibinfo {pages}
  {509} (\bibinfo {year} {2020})}\BibitemShut {NoStop}%
\bibitem [{\citenamefont {Baumgaertel}\ and\ \citenamefont
  {Winter}(1989)}]{baumgaertel89a}%
  \BibitemOpen
  \bibfield  {author} {\bibinfo {author} {\bibfnamefont {M.}~\bibnamefont
  {Baumgaertel}}\ and\ \bibinfo {author} {\bibfnamefont {H.~H.}\ \bibnamefont
  {Winter}},\ }\bibfield  {title} {\bibinfo {title} {Determination of discrete
  relaxation and retardation time spectra from dynamic mechanical data},\
  }\href {https://doi.org/10.1007/BF01332922} {\bibfield  {journal} {\bibinfo
  {journal} {Rheol. Acta}\ }\textbf {\bibinfo {volume} {28}},\ \bibinfo {pages}
  {511} (\bibinfo {year} {1989})}\BibitemShut {NoStop}%
\bibitem [{\citenamefont {Hyun}\ \emph {et~al.}(2007)\citenamefont {Hyun},
  \citenamefont {Baik}, \citenamefont {Ahn}, \citenamefont {Lee}, \citenamefont
  {Sugimoto},\ and\ \citenamefont {Koyama}}]{Hyun2007}%
  \BibitemOpen
  \bibfield  {author} {\bibinfo {author} {\bibfnamefont {K.}~\bibnamefont
  {Hyun}}, \bibinfo {author} {\bibfnamefont {E.~S.}\ \bibnamefont {Baik}},
  \bibinfo {author} {\bibfnamefont {K.~H.}\ \bibnamefont {Ahn}}, \bibinfo
  {author} {\bibfnamefont {S.~J.}\ \bibnamefont {Lee}}, \bibinfo {author}
  {\bibfnamefont {M.}~\bibnamefont {Sugimoto}},\ and\ \bibinfo {author}
  {\bibfnamefont {K.}~\bibnamefont {Koyama}},\ }\bibfield  {title} {\bibinfo
  {title} {Fourier-transform rheology under medium amplitude oscillatory shear
  for linear and branched polymer melts},\ }\href
  {https://doi.org/10.1122/1.2790072} {\bibfield  {journal} {\bibinfo
  {journal} {J. Rheol.}\ }\textbf {\bibinfo {volume} {51}},\ \bibinfo {pages}
  {1319} (\bibinfo {year} {2007})}\BibitemShut {NoStop}%
\bibitem [{\citenamefont {Wagner}\ \emph {et~al.}(2011)\citenamefont {Wagner},
  \citenamefont {Rol\'{o}n-Garrido}, \citenamefont {Hyun},\ and\ \citenamefont
  {Wilhelm}}]{Wagner2011}%
  \BibitemOpen
  \bibfield  {author} {\bibinfo {author} {\bibfnamefont {M.~H.}\ \bibnamefont
  {Wagner}}, \bibinfo {author} {\bibfnamefont {V.~H.}\ \bibnamefont
  {Rol\'{o}n-Garrido}}, \bibinfo {author} {\bibfnamefont {K.}~\bibnamefont
  {Hyun}},\ and\ \bibinfo {author} {\bibfnamefont {M.}~\bibnamefont
  {Wilhelm}},\ }\bibfield  {title} {\bibinfo {title} {Analysis of medium
  amplitude oscillatory shear data of entangled linear and model comb
  polymers},\ }\href {https://doi.org/10.1122/1.3553031} {\bibfield  {journal}
  {\bibinfo  {journal} {J. Rheol.}\ }\textbf {\bibinfo {volume} {55}},\
  \bibinfo {pages} {495} (\bibinfo {year} {2011})}\BibitemShut {NoStop}%
\bibitem [{\citenamefont {Hyun}\ \emph {et~al.}(2011)\citenamefont {Hyun},
  \citenamefont {Wilhelm}, \citenamefont {Klein}, \citenamefont {Cho},
  \citenamefont {Nam}, \citenamefont {Ahn}, \citenamefont {Lee}, \citenamefont
  {Ewoldt},\ and\ \citenamefont {McKinley}}]{Hyun2011}%
  \BibitemOpen
  \bibfield  {author} {\bibinfo {author} {\bibfnamefont {K.}~\bibnamefont
  {Hyun}}, \bibinfo {author} {\bibfnamefont {M.}~\bibnamefont {Wilhelm}},
  \bibinfo {author} {\bibfnamefont {C.~O.}\ \bibnamefont {Klein}}, \bibinfo
  {author} {\bibfnamefont {K.~S.}\ \bibnamefont {Cho}}, \bibinfo {author}
  {\bibfnamefont {J.~G.}\ \bibnamefont {Nam}}, \bibinfo {author} {\bibfnamefont
  {K.~H.}\ \bibnamefont {Ahn}}, \bibinfo {author} {\bibfnamefont {S.~J.}\
  \bibnamefont {Lee}}, \bibinfo {author} {\bibfnamefont {R.~H.}\ \bibnamefont
  {Ewoldt}},\ and\ \bibinfo {author} {\bibfnamefont {G.~H.}\ \bibnamefont
  {McKinley}},\ }\bibfield  {title} {\bibinfo {title} {A review of nonlinear
  oscillatory shear tests: Analysis and application of large amplitude
  oscillatory shear {(LAOS)}},\ }\href
  {https://doi.org/https://doi.org/10.1016/j.progpolymsci.2011.02.002}
  {\bibfield  {journal} {\bibinfo  {journal} {Prog. Polym. Sci.}\ }\textbf
  {\bibinfo {volume} {36}},\ \bibinfo {pages} {1697} (\bibinfo {year}
  {2011})}\BibitemShut {NoStop}%
\bibitem [{\citenamefont {Ewoldt}\ and\ \citenamefont
  {Bharadwaj}(2013)}]{Ewoldt2013}%
  \BibitemOpen
  \bibfield  {author} {\bibinfo {author} {\bibfnamefont {R.~H.}\ \bibnamefont
  {Ewoldt}}\ and\ \bibinfo {author} {\bibfnamefont {N.~A.}\ \bibnamefont
  {Bharadwaj}},\ }\bibfield  {title} {\bibinfo {title} {Low-dimensional
  intrinsic material functions for nonlinear viscoelasticity},\ }\href
  {https://doi.org/10.1007/s00397-013-0686-6} {\bibfield  {journal} {\bibinfo
  {journal} {Rheol. Acta}\ }\textbf {\bibinfo {volume} {52}},\ \bibinfo {pages}
  {201} (\bibinfo {year} {2013})}\BibitemShut {NoStop}%
\bibitem [{\citenamefont {Bharadwaj}\ and\ \citenamefont
  {Ewoldt}(2015)}]{Bharadwaj2015}%
  \BibitemOpen
  \bibfield  {author} {\bibinfo {author} {\bibfnamefont {N.~A.}\ \bibnamefont
  {Bharadwaj}}\ and\ \bibinfo {author} {\bibfnamefont {R.~H.}\ \bibnamefont
  {Ewoldt}},\ }\bibfield  {title} {\bibinfo {title} {Constitutive model
  fingerprints in medium-amplitude oscillatory shear},\ }\href
  {https://doi.org/10.1122/1.4903346} {\bibfield  {journal} {\bibinfo
  {journal} {J. Rheol.}\ }\textbf {\bibinfo {volume} {59}},\ \bibinfo {pages}
  {557} (\bibinfo {year} {2015})}\BibitemShut {NoStop}%
\bibitem [{\citenamefont {Pearson}\ and\ \citenamefont
  {Rochefort}(1982)}]{Pearson1982}%
  \BibitemOpen
  \bibfield  {author} {\bibinfo {author} {\bibfnamefont {D.~S.}\ \bibnamefont
  {Pearson}}\ and\ \bibinfo {author} {\bibfnamefont {W.~E.}\ \bibnamefont
  {Rochefort}},\ }\bibfield  {title} {\bibinfo {title} {Behavior of
  concentrated polystyrene solutions in large-amplitude oscillating shear
  fields},\ }\href {https://doi.org/https://doi.org/10.1002/pol.1982.180200107}
  {\bibfield  {journal} {\bibinfo  {journal} {J. Polym. Sci. Polym. Phys.}\
  }\textbf {\bibinfo {volume} {20}},\ \bibinfo {pages} {83} (\bibinfo {year}
  {1982})}\BibitemShut {NoStop}%
\bibitem [{\citenamefont {Hyun}\ and\ \citenamefont
  {Wilhelm}(2009)}]{Hyun2009}%
  \BibitemOpen
  \bibfield  {author} {\bibinfo {author} {\bibfnamefont {K.}~\bibnamefont
  {Hyun}}\ and\ \bibinfo {author} {\bibfnamefont {M.}~\bibnamefont {Wilhelm}},\
  }\bibfield  {title} {\bibinfo {title} {Establishing a new mechanical
  nonlinear coefficient {Q} from {FT}-rheology: {F}irst investigation of
  entangled linear and comb polymer model systems},\ }\href
  {https://doi.org/10.1021/ma8017266} {\bibfield  {journal} {\bibinfo
  {journal} {Macromolecules}\ }\textbf {\bibinfo {volume} {42}},\ \bibinfo
  {pages} {411} (\bibinfo {year} {2009})}\BibitemShut {NoStop}%
\bibitem [{\citenamefont {Wilhelm}(2002)}]{Wilhelm2002}%
  \BibitemOpen
  \bibfield  {author} {\bibinfo {author} {\bibfnamefont {M.}~\bibnamefont
  {Wilhelm}},\ }\bibfield  {title} {\bibinfo {title} {Fourier-transform
  rheology},\ }\href
  {https://doi.org/https://doi.org/10.1002/1439-2054(20020201)287:2<83::AID-MAME83>3.0.CO;2-B}
  {\bibfield  {journal} {\bibinfo  {journal} {Macromol. Mater. Eng.}\ }\textbf
  {\bibinfo {volume} {287}},\ \bibinfo {pages} {83} (\bibinfo {year}
  {2002})}\BibitemShut {NoStop}%
\bibitem [{\citenamefont {Song}\ \emph {et~al.}(2016)\citenamefont {Song},
  \citenamefont {Nnyigide}, \citenamefont {Salehiyan},\ and\ \citenamefont
  {Hyun}}]{Song2016}%
  \BibitemOpen
  \bibfield  {author} {\bibinfo {author} {\bibfnamefont {H.~Y.}\ \bibnamefont
  {Song}}, \bibinfo {author} {\bibfnamefont {O.~S.}\ \bibnamefont {Nnyigide}},
  \bibinfo {author} {\bibfnamefont {R.}~\bibnamefont {Salehiyan}},\ and\
  \bibinfo {author} {\bibfnamefont {K.}~\bibnamefont {Hyun}},\ }\bibfield
  {title} {\bibinfo {title} {Investigation of nonlinear rheological behavior of
  linear and 3-arm star 1,4-cis-polyisoprene ({PI}) under medium amplitude
  oscillatory shear ({MAOS}) flow via {FT}-rheology},\ }\href
  {https://doi.org/https://doi.org/10.1016/j.polymer.2016.04.052} {\bibfield
  {journal} {\bibinfo  {journal} {Polymer}\ }\textbf {\bibinfo {volume}
  {104}},\ \bibinfo {pages} {268} (\bibinfo {year} {2016})},\ \bibinfo {note}
  {rheology}\BibitemShut {NoStop}%
\bibitem [{\citenamefont {Lee}\ \emph {et~al.}(2016)\citenamefont {Lee},
  \citenamefont {Song},\ and\ \citenamefont {Hyun}}]{Lee2016}%
  \BibitemOpen
  \bibfield  {author} {\bibinfo {author} {\bibfnamefont {S.~H.}\ \bibnamefont
  {Lee}}, \bibinfo {author} {\bibfnamefont {H.~Y.}\ \bibnamefont {Song}},\ and\
  \bibinfo {author} {\bibfnamefont {K.}~\bibnamefont {Hyun}},\ }\bibfield
  {title} {\bibinfo {title} {Effects of silica nanoparticles on copper nanowire
  dispersions in aqueous pva solutions},\ }\href
  {https://doi.org/10.1007/s13367-016-0010-y} {\bibfield  {journal} {\bibinfo
  {journal} {Korea Aust. Rheol. J.}\ }\textbf {\bibinfo {volume} {28}},\
  \bibinfo {pages} {111} (\bibinfo {year} {2016})}\BibitemShut {NoStop}%
\bibitem [{\citenamefont {Lim}\ \emph {et~al.}(2013)\citenamefont {Lim},
  \citenamefont {Ahn}, \citenamefont {Hong},\ and\ \citenamefont
  {Hyun}}]{Lim2013}%
  \BibitemOpen
  \bibfield  {author} {\bibinfo {author} {\bibfnamefont {H.~T.}\ \bibnamefont
  {Lim}}, \bibinfo {author} {\bibfnamefont {K.~H.}\ \bibnamefont {Ahn}},
  \bibinfo {author} {\bibfnamefont {J.~S.}\ \bibnamefont {Hong}},\ and\
  \bibinfo {author} {\bibfnamefont {K.}~\bibnamefont {Hyun}},\ }\bibfield
  {title} {\bibinfo {title} {Nonlinear viscoelasticity of polymer
  nanocomposites under large amplitude oscillatory shear flow},\ }\href
  {https://doi.org/10.1122/1.4795748} {\bibfield  {journal} {\bibinfo
  {journal} {J. Rheol.}\ }\textbf {\bibinfo {volume} {57}},\ \bibinfo {pages}
  {767} (\bibinfo {year} {2013})}\BibitemShut {NoStop}%
\bibitem [{\citenamefont {Ock}\ \emph {et~al.}(2016)\citenamefont {Ock},
  \citenamefont {Ahn}, \citenamefont {Lee},\ and\ \citenamefont
  {Hyun}}]{Ock2016}%
  \BibitemOpen
  \bibfield  {author} {\bibinfo {author} {\bibfnamefont {H.~G.}\ \bibnamefont
  {Ock}}, \bibinfo {author} {\bibfnamefont {K.~H.}\ \bibnamefont {Ahn}},
  \bibinfo {author} {\bibfnamefont {S.~J.}\ \bibnamefont {Lee}},\ and\ \bibinfo
  {author} {\bibfnamefont {K.}~\bibnamefont {Hyun}},\ }\bibfield  {title}
  {\bibinfo {title} {Characterization of compatibilizing effect of organoclay
  in poly(lactic acid) and natural rubber blends by {FT}-rheology},\ }\href
  {https://doi.org/10.1021/acs.macromol.5b02157} {\bibfield  {journal}
  {\bibinfo  {journal} {Macromolecules}\ }\textbf {\bibinfo {volume} {49}},\
  \bibinfo {pages} {2832} (\bibinfo {year} {2016})}\BibitemShut {NoStop}%
\bibitem [{\citenamefont {Salehiyan}\ \emph {et~al.}(2014)\citenamefont
  {Salehiyan}, \citenamefont {Yoo}, \citenamefont {Choi},\ and\ \citenamefont
  {Hyun}}]{Salehiyan2014}%
  \BibitemOpen
  \bibfield  {author} {\bibinfo {author} {\bibfnamefont {R.}~\bibnamefont
  {Salehiyan}}, \bibinfo {author} {\bibfnamefont {Y.}~\bibnamefont {Yoo}},
  \bibinfo {author} {\bibfnamefont {W.~J.}\ \bibnamefont {Choi}},\ and\
  \bibinfo {author} {\bibfnamefont {K.}~\bibnamefont {Hyun}},\ }\bibfield
  {title} {\bibinfo {title} {Characterization of morphologies of compatibilized
  polypropylene/polystyrene blends with nanoparticles via nonlinear rheological
  properties from {FT}-rheology},\ }\href {https://doi.org/10.1021/ma500700e}
  {\bibfield  {journal} {\bibinfo  {journal} {Macromolecules}\ }\textbf
  {\bibinfo {volume} {47}},\ \bibinfo {pages} {4066} (\bibinfo {year}
  {2014})}\BibitemShut {NoStop}%
\bibitem [{\citenamefont {Song}\ and\ \citenamefont {Hyun}(2019)}]{Song2019}%
  \BibitemOpen
  \bibfield  {author} {\bibinfo {author} {\bibfnamefont {H.~Y.}\ \bibnamefont
  {Song}}\ and\ \bibinfo {author} {\bibfnamefont {K.}~\bibnamefont {Hyun}},\
  }\bibfield  {title} {\bibinfo {title} {First-harmonic intrinsic nonlinearity
  of model polymer solutions in medium amplitude oscillatory shear (maos)},\
  }\href {https://doi.org/10.1007/s13367-019-0001-x} {\bibfield  {journal}
  {\bibinfo  {journal} {Korea Aust. Rheol. J.}\ }\textbf {\bibinfo {volume}
  {31}},\ \bibinfo {pages} {1} (\bibinfo {year} {2019})}\BibitemShut {NoStop}%
\bibitem [{\citenamefont {Xiong}\ and\ \citenamefont {Wang}(2018)}]{Xiong2018}%
  \BibitemOpen
  \bibfield  {author} {\bibinfo {author} {\bibfnamefont {W.}~\bibnamefont
  {Xiong}}\ and\ \bibinfo {author} {\bibfnamefont {X.}~\bibnamefont {Wang}},\
  }\bibfield  {title} {\bibinfo {title} {Linear-nonlinear dichotomy of
  rheological responses in particle-filled polymer melts},\ }\href
  {https://doi.org/10.1122/1.4999105} {\bibfield  {journal} {\bibinfo
  {journal} {J. Rheol.}\ }\textbf {\bibinfo {volume} {62}},\ \bibinfo {pages}
  {171} (\bibinfo {year} {2018})}\BibitemShut {NoStop}%
\bibitem [{\citenamefont {Carey-De La~Torre}\ and\ \citenamefont
  {Ewoldt}(2018)}]{Carey-DeLaTorre2018}%
  \BibitemOpen
  \bibfield  {author} {\bibinfo {author} {\bibfnamefont {O.}~\bibnamefont
  {Carey-De La~Torre}}\ and\ \bibinfo {author} {\bibfnamefont {R.~H.}\
  \bibnamefont {Ewoldt}},\ }\bibfield  {title} {\bibinfo {title}
  {First-harmonic nonlinearities can predict unseen third-harmonics in
  medium-amplitude oscillatory shear (maos)},\ }\href
  {https://doi.org/10.1007/s13367-018-0001-2} {\bibfield  {journal} {\bibinfo
  {journal} {Korea-Australia Rheology Journal}\ }\textbf {\bibinfo {volume}
  {30}},\ \bibinfo {pages} {1} (\bibinfo {year} {2018})}\BibitemShut {NoStop}%
\bibitem [{\citenamefont {Hutchings}\ \emph {et~al.}(1992)\citenamefont
  {Hutchings}, \citenamefont {Sheik-Bahae}, \citenamefont {Hagan},\ and\
  \citenamefont {Van~Stryland}}]{Hutchings1992}%
  \BibitemOpen
  \bibfield  {author} {\bibinfo {author} {\bibfnamefont {D.~C.}\ \bibnamefont
  {Hutchings}}, \bibinfo {author} {\bibfnamefont {M.}~\bibnamefont
  {Sheik-Bahae}}, \bibinfo {author} {\bibfnamefont {D.~J.}\ \bibnamefont
  {Hagan}},\ and\ \bibinfo {author} {\bibfnamefont {E.~W.}\ \bibnamefont
  {Van~Stryland}},\ }\bibfield  {title} {\bibinfo {title} {{Kramers-Kr{\"o}nig}
  relations in nonlinear optics},\ }\href {https://doi.org/10.1007/BF01234275}
  {\bibfield  {journal} {\bibinfo  {journal} {Optical and Quantum Electronics}\
  }\textbf {\bibinfo {volume} {24}},\ \bibinfo {pages} {1} (\bibinfo {year}
  {1992})}\BibitemShut {NoStop}%
\bibitem [{\citenamefont {Peiponen}\ \emph {et~al.}(2004)\citenamefont
  {Peiponen}, \citenamefont {Lucarini}, \citenamefont {Saarinen},\ and\
  \citenamefont {Vartiainen}}]{Peiponen2004}%
  \BibitemOpen
  \bibfield  {author} {\bibinfo {author} {\bibfnamefont {K.-E.}\ \bibnamefont
  {Peiponen}}, \bibinfo {author} {\bibfnamefont {V.}~\bibnamefont {Lucarini}},
  \bibinfo {author} {\bibfnamefont {J.~J.}\ \bibnamefont {Saarinen}},\ and\
  \bibinfo {author} {\bibfnamefont {E.}~\bibnamefont {Vartiainen}},\ }\bibfield
   {title} {\bibinfo {title} {{Kramers-Kronig} relations and sum rules in
  nonlinear optical spectroscopy},\ }\href
  {http://www.osapublishing.org/as/abstract.cfm?URI=as-58-5- 499} {\bibfield
  {journal} {\bibinfo  {journal} {Appl. Spectrosc.}\ }\textbf {\bibinfo
  {volume} {58}},\ \bibinfo {pages} {499} (\bibinfo {year} {2004})}\BibitemShut
  {NoStop}%
\bibitem [{\citenamefont {Boyd}(2008)}]{Boyd2008}%
  \BibitemOpen
  \bibfield  {author} {\bibinfo {author} {\bibfnamefont {R.~W.}\ \bibnamefont
  {Boyd}},\ }\bibfield  {title} {\bibinfo {title} {Chapter 1: The nonlinear
  optical susceptibility},\ }in\ \href
  {https://doi.org/https://doi.org/10.1016/B978-0-12-369470-6.00001-0} {\emph
  {\bibinfo {booktitle} {Nonlinear Optics}}},\ \bibinfo {editor} {edited by\
  \bibinfo {editor} {\bibfnamefont {R.~W.}\ \bibnamefont {Boyd}}}\ (\bibinfo
  {publisher} {Academic Press},\ \bibinfo {address} {Burlington},\ \bibinfo
  {year} {2008})\ \bibinfo {edition} {third edition}\ ed.,\ pp.\ \bibinfo
  {pages} {1--67}\BibitemShut {NoStop}%
\bibitem [{\citenamefont {Cho}\ \emph {et~al.}(2010)\citenamefont {Cho},
  \citenamefont {Song},\ and\ \citenamefont {Chang}}]{Cho2010}%
  \BibitemOpen
  \bibfield  {author} {\bibinfo {author} {\bibfnamefont {K.~S.}\ \bibnamefont
  {Cho}}, \bibinfo {author} {\bibfnamefont {K.-W.}\ \bibnamefont {Song}},\ and\
  \bibinfo {author} {\bibfnamefont {G.-S.}\ \bibnamefont {Chang}},\ }\bibfield
  {title} {\bibinfo {title} {Scaling relations in nonlinear viscoelastic
  behavior of aqueous {PEO} solutions under large amplitude oscillatory shear
  flow},\ }\href {https://doi.org/10.1122/1.3258278} {\bibfield  {journal}
  {\bibinfo  {journal} {J. Rheol.}\ }\textbf {\bibinfo {volume} {54}},\
  \bibinfo {pages} {27} (\bibinfo {year} {2010})}\BibitemShut {NoStop}%
\bibitem [{\citenamefont {Martinetti}\ and\ \citenamefont
  {Ewoldt}(2019)}]{Martinetti2019}%
  \BibitemOpen
  \bibfield  {author} {\bibinfo {author} {\bibfnamefont {L.}~\bibnamefont
  {Martinetti}}\ and\ \bibinfo {author} {\bibfnamefont {R.~H.}\ \bibnamefont
  {Ewoldt}},\ }\bibfield  {title} {\bibinfo {title} {Time-strain separability
  in medium-amplitude oscillatory shear},\ }\href
  {https://doi.org/10.1063/1.5085025} {\bibfield  {journal} {\bibinfo
  {journal} {Phys. Fluids}\ }\textbf {\bibinfo {volume} {31}},\ \bibinfo
  {pages} {021213} (\bibinfo {year} {2019})}\BibitemShut {NoStop}%
\bibitem [{\citenamefont {Lennon}\ \emph {et~al.}(2020)\citenamefont {Lennon},
  \citenamefont {McKinley},\ and\ \citenamefont {Swan}}]{Lennon2020}%
  \BibitemOpen
  \bibfield  {author} {\bibinfo {author} {\bibfnamefont {K.~R.}\ \bibnamefont
  {Lennon}}, \bibinfo {author} {\bibfnamefont {G.~H.}\ \bibnamefont
  {McKinley}},\ and\ \bibinfo {author} {\bibfnamefont {J.~W.}\ \bibnamefont
  {Swan}},\ }\bibfield  {title} {\bibinfo {title} {Medium amplitude parallel
  superposition ({MAPS}) rheology. {Part 1: Mathematical} framework and
  theoretical examples},\ }\href {https://doi.org/10.1122/1.5132693} {\bibfield
   {journal} {\bibinfo  {journal} {J. Rheol.}\ }\textbf {\bibinfo {volume}
  {64}},\ \bibinfo {pages} {551} (\bibinfo {year} {2020})}\BibitemShut
  {NoStop}%
\bibitem [{\citenamefont {Liu}\ \emph {et~al.}(2020)\citenamefont {Liu},
  \citenamefont {Xiong}, \citenamefont {Nie},\ and\ \citenamefont
  {Yu}}]{Liu2020}%
  \BibitemOpen
  \bibfield  {author} {\bibinfo {author} {\bibfnamefont {Z.}~\bibnamefont
  {Liu}}, \bibinfo {author} {\bibfnamefont {Z.}~\bibnamefont {Xiong}}, \bibinfo
  {author} {\bibfnamefont {Z.}~\bibnamefont {Nie}},\ and\ \bibinfo {author}
  {\bibfnamefont {W.}~\bibnamefont {Yu}},\ }\bibfield  {title} {\bibinfo
  {title} {Correlation between linear and nonlinear material functions under
  large amplitude oscillatory shear},\ }\href
  {https://doi.org/10.1063/5.0021792} {\bibfield  {journal} {\bibinfo
  {journal} {Phys. Fluids}\ }\textbf {\bibinfo {volume} {32}},\ \bibinfo
  {pages} {093105} (\bibinfo {year} {2020})}\BibitemShut {NoStop}%
\bibitem [{\citenamefont {Singh}\ \emph {et~al.}(2018)\citenamefont {Singh},
  \citenamefont {Soulages},\ and\ \citenamefont {Ewoldt}}]{Singh2018}%
  \BibitemOpen
  \bibfield  {author} {\bibinfo {author} {\bibfnamefont {P.~K.}\ \bibnamefont
  {Singh}}, \bibinfo {author} {\bibfnamefont {J.~M.}\ \bibnamefont
  {Soulages}},\ and\ \bibinfo {author} {\bibfnamefont {R.~H.}\ \bibnamefont
  {Ewoldt}},\ }\bibfield  {title} {\bibinfo {title} {Frequency-sweep
  medium-amplitude oscillatory shear (maos)},\ }\href
  {https://doi.org/10.1122/1.4999795} {\bibfield  {journal} {\bibinfo
  {journal} {Journal of Rheology}\ }\textbf {\bibinfo {volume} {62}},\ \bibinfo
  {pages} {277} (\bibinfo {year} {2018})}\BibitemShut {NoStop}%
\bibitem [{\citenamefont {Tibshirani}(1996)}]{Tibshirani1996}%
  \BibitemOpen
  \bibfield  {author} {\bibinfo {author} {\bibfnamefont {R.}~\bibnamefont
  {Tibshirani}},\ }\bibfield  {title} {\bibinfo {title} {Regression shrinkage
  and selection via the lasso},\ }\href
  {https://doi.org/https://doi.org/10.1111/j.2517-6161.1996.tb02080.x}
  {\bibfield  {journal} {\bibinfo  {journal} {J. R. Stat. Soc. Series B Stat.
  Methodol.}\ }\textbf {\bibinfo {volume} {58}},\ \bibinfo {pages} {267}
  (\bibinfo {year} {1996})}\BibitemShut {NoStop}%
\bibitem [{\citenamefont {Tibshirani}(2011)}]{Tibshirani2011}%
  \BibitemOpen
  \bibfield  {author} {\bibinfo {author} {\bibfnamefont {R.}~\bibnamefont
  {Tibshirani}},\ }\bibfield  {title} {\bibinfo {title} {Regression shrinkage
  and selection via the lasso: a retrospective},\ }\href
  {https://doi.org/https://doi.org/10.1111/j.1467-9868.2011.00771.x} {\bibfield
   {journal} {\bibinfo  {journal} {J. R. Stat. Soc. Series B Stat. Methodol.}\
  }\textbf {\bibinfo {volume} {73}},\ \bibinfo {pages} {273} (\bibinfo {year}
  {2011})}\BibitemShut {NoStop}%
\bibitem [{\citenamefont {Bharadwaj}(2016)}]{bharadwaj2016dissertation}%
  \BibitemOpen
  \bibfield  {author} {\bibinfo {author} {\bibfnamefont {N.~A.~K.}\
  \bibnamefont {Bharadwaj}},\ }\emph {\bibinfo {title} {Asymptotically
  nonlinear oscillatory shear: {Theory}, modeling, measurements and
  applications of nonlinear elasticity to stimuli-responsive composites}},\
  \href@noop {} {Ph.D. thesis},\ \bibinfo  {school} {University of Illinois at
  Urbana-Champaign} (\bibinfo {year} {2016})\BibitemShut {NoStop}%
\bibitem [{\citenamefont {Pedregosa}\ \emph {et~al.}(2011)\citenamefont
  {Pedregosa}, \citenamefont {Varoquaux}, \citenamefont {Gramfort},
  \citenamefont {Michel}, \citenamefont {Thirion}, \citenamefont {Grisel},
  \citenamefont {Blondel}, \citenamefont {Prettenhofer}, \citenamefont {Weiss},
  \citenamefont {Dubourg}, \citenamefont {Vanderplas}, \citenamefont {Passos},
  \citenamefont {Cournapeau}, \citenamefont {Brucher}, \citenamefont {Perrot},\
  and\ \citenamefont {Duchesnay}}]{Pedregosa2011}%
  \BibitemOpen
  \bibfield  {author} {\bibinfo {author} {\bibfnamefont {F.}~\bibnamefont
  {Pedregosa}}, \bibinfo {author} {\bibfnamefont {G.}~\bibnamefont
  {Varoquaux}}, \bibinfo {author} {\bibfnamefont {A.}~\bibnamefont {Gramfort}},
  \bibinfo {author} {\bibfnamefont {V.}~\bibnamefont {Michel}}, \bibinfo
  {author} {\bibfnamefont {B.}~\bibnamefont {Thirion}}, \bibinfo {author}
  {\bibfnamefont {O.}~\bibnamefont {Grisel}}, \bibinfo {author} {\bibfnamefont
  {M.}~\bibnamefont {Blondel}}, \bibinfo {author} {\bibfnamefont
  {P.}~\bibnamefont {Prettenhofer}}, \bibinfo {author} {\bibfnamefont
  {R.}~\bibnamefont {Weiss}}, \bibinfo {author} {\bibfnamefont
  {V.}~\bibnamefont {Dubourg}}, \bibinfo {author} {\bibfnamefont
  {J.}~\bibnamefont {Vanderplas}}, \bibinfo {author} {\bibfnamefont
  {A.}~\bibnamefont {Passos}}, \bibinfo {author} {\bibfnamefont
  {D.}~\bibnamefont {Cournapeau}}, \bibinfo {author} {\bibfnamefont
  {M.}~\bibnamefont {Brucher}}, \bibinfo {author} {\bibfnamefont
  {M.}~\bibnamefont {Perrot}},\ and\ \bibinfo {author} {\bibfnamefont
  {E.}~\bibnamefont {Duchesnay}},\ }\bibfield  {title} {\bibinfo {title}
  {Scikit-learn: {M}achine learning in {P}ython},\ }\href@noop {} {\bibfield
  {journal} {\bibinfo  {journal} {J. Mach. Learn. Res.}\ }\textbf {\bibinfo
  {volume} {12}},\ \bibinfo {pages} {2825} (\bibinfo {year}
  {2011})}\BibitemShut {NoStop}%
\bibitem [{\citenamefont {Friedman}\ \emph {et~al.}(2010)\citenamefont
  {Friedman}, \citenamefont {Hastie},\ and\ \citenamefont
  {Tibshirani}}]{Friedman2010}%
  \BibitemOpen
  \bibfield  {author} {\bibinfo {author} {\bibfnamefont {J.}~\bibnamefont
  {Friedman}}, \bibinfo {author} {\bibfnamefont {T.}~\bibnamefont {Hastie}},\
  and\ \bibinfo {author} {\bibfnamefont {R.}~\bibnamefont {Tibshirani}},\
  }\bibfield  {title} {\bibinfo {title} {Regularization paths for generalized
  linear models via coordinate descent},\ }\href@noop {} {\bibfield  {journal}
  {\bibinfo  {journal} {J. Stat. Softw.}\ }\textbf {\bibinfo {volume} {33}},\
  \bibinfo {pages} {1} (\bibinfo {year} {2010})},\ \bibinfo {note}
  {20808728[pmid]}\BibitemShut {NoStop}%
\bibitem [{\citenamefont {Kim}\ \emph {et~al.}(2008)\citenamefont {Kim},
  \citenamefont {Koh}, \citenamefont {Lustig}, \citenamefont {Boyd},\ and\
  \citenamefont {Gorinevsky}}]{Kim2008}%
  \BibitemOpen
  \bibfield  {author} {\bibinfo {author} {\bibfnamefont {S.}~\bibnamefont
  {Kim}}, \bibinfo {author} {\bibfnamefont {K.}~\bibnamefont {Koh}}, \bibinfo
  {author} {\bibfnamefont {M.}~\bibnamefont {Lustig}}, \bibinfo {author}
  {\bibfnamefont {S.}~\bibnamefont {Boyd}},\ and\ \bibinfo {author}
  {\bibfnamefont {D.}~\bibnamefont {Gorinevsky}},\ }\bibfield  {title}
  {\bibinfo {title} {An interior-point method for large-scale l1-regularized
  least squares},\ }\href@noop {} {\bibfield  {journal} {\bibinfo  {journal}
  {IEEE J. Sel. Top. Signal Process.}\ }\textbf {\bibinfo {volume} {1}},\
  \bibinfo {pages} {606} (\bibinfo {year} {2008})}\BibitemShut {NoStop}%
\bibitem [{\citenamefont {Giesekus}(1982)}]{Giesekus1982}%
  \BibitemOpen
  \bibfield  {author} {\bibinfo {author} {\bibfnamefont {H.}~\bibnamefont
  {Giesekus}},\ }\bibfield  {title} {\bibinfo {title} {A simple constitutive
  equation for polymer fluids based on the concept of deformation-dependent
  tensorial mobility},\ }\href
  {https://doi.org/https://doi.org/10.1016/0377-0257(82)85016-7} {\bibfield
  {journal} {\bibinfo  {journal} {J. Non-Newtonian Fluid Mech.}\ }\textbf
  {\bibinfo {volume} {11}},\ \bibinfo {pages} {69} (\bibinfo {year}
  {1982})}\BibitemShut {NoStop}%
\bibitem [{\citenamefont {Larson}(1998)}]{larsoncf}%
  \BibitemOpen
  \bibfield  {author} {\bibinfo {author} {\bibfnamefont {R.~G.}\ \bibnamefont
  {Larson}},\ }\href@noop {} {\emph {\bibinfo {title} {{Structure and Rheology
  of Complex Fluids}}}}\ (\bibinfo  {publisher} {Oxford University Press},\
  \bibinfo {address} {New York},\ \bibinfo {year} {1998})\BibitemShut {NoStop}%
\bibitem [{\citenamefont {Fischer}\ and\ \citenamefont
  {Rehage}(1997)}]{Fischer1997}%
  \BibitemOpen
  \bibfield  {author} {\bibinfo {author} {\bibfnamefont {P.}~\bibnamefont
  {Fischer}}\ and\ \bibinfo {author} {\bibfnamefont {H.}~\bibnamefont
  {Rehage}},\ }\bibfield  {title} {\bibinfo {title} {Non-linear flow properties
  of viscoelastic surfactant solutions},\ }\href
  {https://doi.org/10.1007/BF00366720} {\bibfield  {journal} {\bibinfo
  {journal} {Rheol. Acta}\ }\textbf {\bibinfo {volume} {36}},\ \bibinfo {pages}
  {13} (\bibinfo {year} {1997})}\BibitemShut {NoStop}%
\bibitem [{\citenamefont {Helgeson}\ \emph {et~al.}(2010)\citenamefont
  {Helgeson}, \citenamefont {Hodgdon}, \citenamefont {Kaler},\ and\
  \citenamefont {Wagner}}]{Helgeson2010}%
  \BibitemOpen
  \bibfield  {author} {\bibinfo {author} {\bibfnamefont {M.~E.}\ \bibnamefont
  {Helgeson}}, \bibinfo {author} {\bibfnamefont {T.~K.}\ \bibnamefont
  {Hodgdon}}, \bibinfo {author} {\bibfnamefont {E.~W.}\ \bibnamefont {Kaler}},\
  and\ \bibinfo {author} {\bibfnamefont {N.~J.}\ \bibnamefont {Wagner}},\
  }\bibfield  {title} {\bibinfo {title} {A systematic study of equilibrium
  structure, thermodynamics, and rheology of aqueous {CTAB/NaNO$_3$} wormlike
  micelles},\ }\href
  {https://doi.org/https://doi.org/10.1016/j.jcis.2010.05.045} {\bibfield
  {journal} {\bibinfo  {journal} {J. Colloid Interface Sci.}\ }\textbf
  {\bibinfo {volume} {349}},\ \bibinfo {pages} {1} (\bibinfo {year}
  {2010})}\BibitemShut {NoStop}%
\bibitem [{\citenamefont {Kate~Gurnon}\ and\ \citenamefont
  {Wagner}(2012)}]{KateGurnon2012}%
  \BibitemOpen
  \bibfield  {author} {\bibinfo {author} {\bibfnamefont {A.}~\bibnamefont
  {Kate~Gurnon}}\ and\ \bibinfo {author} {\bibfnamefont {N.~J.}\ \bibnamefont
  {Wagner}},\ }\bibfield  {title} {\bibinfo {title} {Large amplitude
  oscillatory shear {(LAOS)} measurements to obtain constitutive equation model
  parameters: {Giesekus} model of banding and nonbanding wormlike micelles},\
  }\href {https://doi.org/10.1122/1.3684751} {\bibfield  {journal} {\bibinfo
  {journal} {J. Rheol.}\ }\textbf {\bibinfo {volume} {56}},\ \bibinfo {pages}
  {333} (\bibinfo {year} {2012})}\BibitemShut {NoStop}%
\bibitem [{\citenamefont {Holz}\ \emph {et~al.}(1999)\citenamefont {Holz},
  \citenamefont {Fischer},\ and\ \citenamefont {Rehage}}]{Holz1999}%
  \BibitemOpen
  \bibfield  {author} {\bibinfo {author} {\bibfnamefont {T.}~\bibnamefont
  {Holz}}, \bibinfo {author} {\bibfnamefont {P.}~\bibnamefont {Fischer}},\ and\
  \bibinfo {author} {\bibfnamefont {H.}~\bibnamefont {Rehage}},\ }\bibfield
  {title} {\bibinfo {title} {Shear relaxation in the nonlinear-viscoelastic
  regime of a giesekus fluid},\ }\href
  {https://doi.org/https://doi.org/10.1016/S0377-0257(99)00016-6} {\bibfield
  {journal} {\bibinfo  {journal} {J. Non-Newtonian Fluid Mech.}\ }\textbf
  {\bibinfo {volume} {88}},\ \bibinfo {pages} {133} (\bibinfo {year}
  {1999})}\BibitemShut {NoStop}%
\bibitem [{\citenamefont {Larson}(1985)}]{Larson1985}%
  \BibitemOpen
  \bibfield  {author} {\bibinfo {author} {\bibfnamefont {R.}~\bibnamefont
  {Larson}},\ }\bibfield  {title} {\bibinfo {title} {Constitutive relationships
  for polymeric materials with power-law distributions of relaxation times},\
  }\href {https://doi.org/10.1007/BF01333961} {\bibfield  {journal} {\bibinfo
  {journal} {Rheol. Acta}\ }\textbf {\bibinfo {volume} {24}},\ \bibinfo {pages}
  {327} (\bibinfo {year} {1985})}\BibitemShut {NoStop}%
\bibitem [{\citenamefont {Campanella}\ and\ \citenamefont
  {Peleg}(1987)}]{Campanella1987}%
  \BibitemOpen
  \bibfield  {author} {\bibinfo {author} {\bibfnamefont {O.}~\bibnamefont
  {Campanella}}\ and\ \bibinfo {author} {\bibfnamefont {M.}~\bibnamefont
  {Peleg}},\ }\bibfield  {title} {\bibinfo {title} {Analysis of the transient
  flow of mayonnaise in a coaxial viscometer},\ }\href
  {https://doi.org/10.1122/1.549931} {\bibfield  {journal} {\bibinfo  {journal}
  {J. Rheol.}\ }\textbf {\bibinfo {volume} {31}},\ \bibinfo {pages} {439}
  (\bibinfo {year} {1987})}\BibitemShut {NoStop}%
\bibitem [{\citenamefont {Weir}\ \emph {et~al.}(2016)\citenamefont {Weir},
  \citenamefont {Bromley}, \citenamefont {Lips},\ and\ \citenamefont
  {Poon}}]{Poon2016}%
  \BibitemOpen
  \bibfield  {author} {\bibinfo {author} {\bibfnamefont {S.}~\bibnamefont
  {Weir}}, \bibinfo {author} {\bibfnamefont {K.}~\bibnamefont {Bromley}},
  \bibinfo {author} {\bibfnamefont {A.}~\bibnamefont {Lips}},\ and\ \bibinfo
  {author} {\bibfnamefont {W.}~\bibnamefont {Poon}},\ }\bibfield  {title}
  {\bibinfo {title} {{Celebrating Soft Matter's 10$^\text{th}$ Anniversary:
  Simplicity in complexity - Towards a soft matter physics of caramel}},\
  }\href {https://doi.org/10.1039/c5sm01797a} {\bibfield  {journal} {\bibinfo
  {journal} {Soft Matter}\ }\textbf {\bibinfo {volume} {12}},\ \bibinfo {pages}
  {2757} (\bibinfo {year} {2016})}\BibitemShut {NoStop}%
\bibitem [{\citenamefont {Rathinaraj}\ \emph {et~al.}(2021)\citenamefont
  {Rathinaraj}, \citenamefont {McKinley},\ and\ \citenamefont
  {Keshavarz}}]{Rathinaraj2021}%
  \BibitemOpen
  \bibfield  {author} {\bibinfo {author} {\bibfnamefont {J.~D.~J.}\
  \bibnamefont {Rathinaraj}}, \bibinfo {author} {\bibfnamefont {G.~H.}\
  \bibnamefont {McKinley}},\ and\ \bibinfo {author} {\bibfnamefont
  {B.}~\bibnamefont {Keshavarz}},\ }\bibfield  {title} {\bibinfo {title}
  {Incorporating rheological nonlinearity into fractional calculus descriptions
  of fractal matter and multi-scale complex fluids},\ }\bibfield  {journal}
  {\bibinfo  {journal} {Fractal and Fractional}\ }\textbf {\bibinfo {volume}
  {5}},\ \href {https://doi.org/10.3390/fractalfract5040174}
  {10.3390/fractalfract5040174} (\bibinfo {year} {2021})\BibitemShut {NoStop}%
\bibitem [{\citenamefont {Suman}\ \emph {et~al.}(2021)\citenamefont {Suman},
  \citenamefont {Shanbhag},\ and\ \citenamefont {Joshi}}]{Suman2021}%
  \BibitemOpen
  \bibfield  {author} {\bibinfo {author} {\bibfnamefont {K.}~\bibnamefont
  {Suman}}, \bibinfo {author} {\bibfnamefont {S.}~\bibnamefont {Shanbhag}},\
  and\ \bibinfo {author} {\bibfnamefont {Y.~M.}\ \bibnamefont {Joshi}},\
  }\bibfield  {title} {\bibinfo {title} {Phenomenological model of
  viscoelasticity for systems undergoing sol-gel transition},\ }\href
  {https://doi.org/10.1063/5.0038830} {\bibfield  {journal} {\bibinfo
  {journal} {Phys. Fluids}\ }\textbf {\bibinfo {volume} {33}},\ \bibinfo
  {pages} {033103} (\bibinfo {year} {2021})}\BibitemShut {NoStop}%
\bibitem [{\citenamefont {Suman}\ and\ \citenamefont
  {Joshi}(2020)}]{Suman2020}%
  \BibitemOpen
  \bibfield  {author} {\bibinfo {author} {\bibfnamefont {K.}~\bibnamefont
  {Suman}}\ and\ \bibinfo {author} {\bibfnamefont {Y.~M.}\ \bibnamefont
  {Joshi}},\ }\bibfield  {title} {\bibinfo {title} {On the universality of the
  scaling relations during sol-gel transition},\ }\href
  {https://doi.org/10.1122/1.5134115} {\bibfield  {journal} {\bibinfo
  {journal} {J. Rheol.}\ }\textbf {\bibinfo {volume} {64}},\ \bibinfo {pages}
  {863} (\bibinfo {year} {2020})}\BibitemShut {NoStop}%
\bibitem [{\citenamefont {Keshavarz}\ \emph {et~al.}(2017)\citenamefont
  {Keshavarz}, \citenamefont {Divoux}, \citenamefont {Manneville},\ and\
  \citenamefont {McKinley}}]{Bavand2017}%
  \BibitemOpen
  \bibfield  {author} {\bibinfo {author} {\bibfnamefont {B.}~\bibnamefont
  {Keshavarz}}, \bibinfo {author} {\bibfnamefont {T.}~\bibnamefont {Divoux}},
  \bibinfo {author} {\bibfnamefont {S.}~\bibnamefont {Manneville}},\ and\
  \bibinfo {author} {\bibfnamefont {G.~H.}\ \bibnamefont {McKinley}},\
  }\bibfield  {title} {\bibinfo {title} {Nonlinear viscoelasticity and
  generalized failure criterion for polymer gels},\ }\href
  {https://doi.org/10.1021/acsmacrolett.7b00213} {\bibfield  {journal}
  {\bibinfo  {journal} {ACS Macro Letters}\ }\textbf {\bibinfo {volume} {6}},\
  \bibinfo {pages} {663} (\bibinfo {year} {2017})}\BibitemShut {NoStop}%
\bibitem [{\citenamefont {Suman}\ and\ \citenamefont
  {Joshi}(2019)}]{Suman2019}%
  \BibitemOpen
  \bibfield  {author} {\bibinfo {author} {\bibfnamefont {K.}~\bibnamefont
  {Suman}}\ and\ \bibinfo {author} {\bibfnamefont {Y.~M.}\ \bibnamefont
  {Joshi}},\ }\bibfield  {title} {\bibinfo {title} {Analyzing onset of
  nonlinearity of a colloidal gel at the critical point},\ }\href
  {https://doi.org/10.1122/1.5108611} {\bibfield  {journal} {\bibinfo
  {journal} {J. Rheol.}\ }\textbf {\bibinfo {volume} {63}},\ \bibinfo {pages}
  {991} (\bibinfo {year} {2019})}\BibitemShut {NoStop}%
\bibitem [{\citenamefont {Shanbhag}\ \emph {et~al.}(2021)\citenamefont
  {Shanbhag}, \citenamefont {Mittal},\ and\ \citenamefont
  {Joshi}}]{Shanbhag2021}%
  \BibitemOpen
  \bibfield  {author} {\bibinfo {author} {\bibfnamefont {S.}~\bibnamefont
  {Shanbhag}}, \bibinfo {author} {\bibfnamefont {S.}~\bibnamefont {Mittal}},\
  and\ \bibinfo {author} {\bibfnamefont {Y.~M.}\ \bibnamefont {Joshi}},\
  }\bibfield  {title} {\bibinfo {title} {Spectral method for time-strain
  separable integral constitutive models in oscillatory shear},\ }\href
  {https://doi.org/10.1063/5.0072377} {\bibfield  {journal} {\bibinfo
  {journal} {Phys. Fluids}\ }\textbf {\bibinfo {volume} {33}},\ \bibinfo
  {pages} {113104} (\bibinfo {year} {2021})}\BibitemShut {NoStop}%
\bibitem [{\citenamefont {Davis}\ and\ \citenamefont
  {Macosko}(1978)}]{Davis1978}%
  \BibitemOpen
  \bibfield  {author} {\bibinfo {author} {\bibfnamefont {W.~M.}\ \bibnamefont
  {Davis}}\ and\ \bibinfo {author} {\bibfnamefont {C.~W.}\ \bibnamefont
  {Macosko}},\ }\bibfield  {title} {\bibinfo {title} {Nonlinear dynamic
  mechanical moduli for polycarbonate and pmma},\ }\href
  {https://doi.org/10.1122/1.549500} {\bibfield  {journal} {\bibinfo  {journal}
  {J. Rheol.}\ }\textbf {\bibinfo {volume} {22}},\ \bibinfo {pages} {53}
  (\bibinfo {year} {1978})}\BibitemShut {NoStop}%
\bibitem [{\citenamefont {Bharadwaj}\ \emph {et~al.}(2017)\citenamefont
  {Bharadwaj}, \citenamefont {Schweizer},\ and\ \citenamefont
  {Ewoldt}}]{Bharadwaj2017}%
  \BibitemOpen
  \bibfield  {author} {\bibinfo {author} {\bibfnamefont {N.~A.}\ \bibnamefont
  {Bharadwaj}}, \bibinfo {author} {\bibfnamefont {K.~S.}\ \bibnamefont
  {Schweizer}},\ and\ \bibinfo {author} {\bibfnamefont {R.~H.}\ \bibnamefont
  {Ewoldt}},\ }\bibfield  {title} {\bibinfo {title} {A strain stiffening theory
  for transient polymer networks under asymptotically nonlinear oscillatory
  shear},\ }\href {https://doi.org/10.1122/1.4979368} {\bibfield  {journal}
  {\bibinfo  {journal} {J. Rheol.}\ }\textbf {\bibinfo {volume} {61}},\
  \bibinfo {pages} {643} (\bibinfo {year} {2017})}\BibitemShut {NoStop}%
\bibitem [{\citenamefont {Davies}\ and\ \citenamefont
  {Anderssen}(1997)}]{Davies1997}%
  \BibitemOpen
  \bibfield  {author} {\bibinfo {author} {\bibfnamefont {A.}~\bibnamefont
  {Davies}}\ and\ \bibinfo {author} {\bibfnamefont {R.}~\bibnamefont
  {Anderssen}},\ }\bibfield  {title} {\bibinfo {title} {Sampling localization
  in determining the relaxation spectrum},\ }\href
  {https://doi.org/https://doi.org/10.1016/S0377-0257(97)00056-6} {\bibfield
  {journal} {\bibinfo  {journal} {J. Non-Newtonian Fluid Mech.}\ }\textbf
  {\bibinfo {volume} {73}},\ \bibinfo {pages} {163} (\bibinfo {year}
  {1997})}\BibitemShut {NoStop}%
\end{thebibliography}%

\end{document}


\begin{center}
\textbf{\huge{Supplementary Information}}\\
\vspace{0.5cm}
\textbf{\Large Kramers-Kronig Relations for Nonlinear Rheology: 2. Validation of Medium Amplitude Oscillatory Shear (MAOS) Measurements}\\
\vspace{0.5cm}

\medskip 

Sachin Shanbhag\footnote{sshanbhag@fsu.edu}\\
\textit{Department of Scientific Computing}\\
\textit{Florida State University, Tallahassee, FL 32306, USA}\\

\medskip
Yogesh M. Joshi\footnote{joshi@iitk.ac.in}\\
\textit{Department of Chemical Engineering,}\\
\textit{Indian Institute of Technology, Kanpur, INDIA.}

\vspace{0.5cm}
\end{center}

\tableofcontents

\newpage

\section{Python Code for SMEL Test}

This code was tested on a workstation running Debian Linux (Linux Mint 19.3 Tricia), Python (3.6.9), NumPy (1.19.5), and scikit-learn (0.23.2).

\begin{minted}[fontsize=\small]{python}

import numpy as np
from sklearn.linear_model import LassoCV

def lassoFit(wexp, Gexp, decade_density=10):
    """
    Inputs:
        wexp: array of size n containing the experimental frequency
        Gexp: array of size 2n containing G33' stacked over G33" 
        decade_density: density of modes per decade; 
                        the initial number of modes N is inferred
        
    Outputs:
        g, tau : arrays of length <N for Maxwell modes.
                 Only modes with nonzero gj are returned
        alpha  : the optimal value of alpha used in LASSO
        score  : coefficient of determination R2; 
                 used to quantitatively assess quality of fit

    """

    # odd number of modes, puts a mode at the center of data range
    wmin = min(wexp); wmax = max(wexp)
    N    = decade_density * int(np.log10(wmax/wmin) + 2)
    tau = np.geomspace(0.1/wmax, 10/wmin, N)

    # Set up regression problem so that Gexp = K * g, 
    # where K is 2n*N, g = N*1, and Gst = 2n*1
    S, W    = np.meshgrid(tau, wexp)
    ws      = S*W
    ws2     = ws**2
    K1      = 3 * ws2/(1+ws2) - 3 * (4*ws2/(1+4*ws2)) + 1 * (9*ws2/(1+9*ws2))
    K2      = 3 * ws/(1+ws2)  - 3 * (2*ws/(1+4*ws2)) + 1 * (3*ws/(1+9*ws2))
    K       = np.vstack((K1, K2))   # 2n * N

    # actual fitting
    WtMat   = np.diag(np.sqrt(1./np.abs(Gexp))) # 2n*2n diagonal matrix
    K_trans = np.dot(WtMat, K)
    G_trans = np.dot(WtMat, Gexp)
    model   = LassoCV(cv=3, fit_intercept=False)
    
    model.fit(K_trans, G_trans)

    score = model.score(K_trans, G_trans)  # R2
    alpha = model.alpha_
    cnd   = np.abs(model.coef_) > 1.0e-16
    g     = model.coef_[cnd]
    tau   = tau[cnd]
        
    return g, tau, alpha, score
\end{minted}

\clearpage
\newpage